\documentclass{article}
\usepackage{graphicx} 
\usepackage{jcappub}
\usepackage[dvipsnames]{xcolor}
\usepackage[normalem]{ulem}
\usepackage[noabbrev]{cleveref}
\usepackage{csquotes}
\usepackage{orcidlink}
\usepackage{hyperref}

\title{The BAO scale -- how standard is the standard ruler?}
\author[1,2]{{Francisco Asensio-Rivera}\orcidlink{0009-0008-4207-5955},}
\author[3,4]{{Nils Sch\"oneberg}\orcidlink{0000-0002-7873-0404},}
\author[1,2,5]{{H\'ector Gil-Mar\'in}\orcidlink{0000-0003-0265-6217},}
\author[1,6]{{Licia Verde}\orcidlink{0000-0003-2601-8770}}
\emailAdd{franasensio@icc.ub.edu }
\emailAdd{N.Schoeneberg@lmu.de}
\emailAdd{hectorgil@icc.ub.edu}
\emailAdd{liciaverde@icc.ub.edu}

\affiliation[1]{Institut de Ci\`encies del Cosmos (ICCUB), Universitat de Barcelona (UB), c. Mart\'i i Franqu\`es, 1, 08028 Barcelona, Spain}

\affiliation[2]{Departament de F\'{\i}sica Qu\`{a}ntica i Astrof\'{\i}sica (FQA), Universitat de Barcelona (UB), c. Mart\'{\i} i Franqu\`{e}s, 1, 08028 Barcelona, Spain}

\affiliation[3]{University Observatory, Faculty of Physics, Ludwig-Maximilians-Universität, Scheinerstr. 1, 81677 München, Germany}

\affiliation[4]{Excellence Cluster ORIGINS, Boltzmannstrasse 2, D-85748 Garching, Germany}

\affiliation[5]{Institut d'Estudis Espacials de Catalunya (IEEC), c/ Esteve Terradas 1, Edifici RDIT, Campus PMT-UPC, 08860 Castelldefels, Spain}

\affiliation[6]{Instituci\'{o} Catalana de Recerca i Estudis Avan\c{c}ats, Passeig de Llu\'{\i}s Companys, 23, 08010 Barcelona, Spain}

\abstract{
Analyses of baryon acoustic oscillations (BAO) commonly employ template-based methods to extract compressed parameters from the clustering of dark-matter tracers, which are then interpreted in terms of ratios of the sound-horizon scale and cosmological distances relative to a fiducial cosmology. A small mismatch between the sound-horizon scale derived from the standard analytic formulation (integral over the sound speed) and the effective scale imprinted in clustered matter can, however, introduce a systematic bias in cosmological inference.
We extend previous work to a broader class of cosmological models, quantify this bias for surveys with DESI-like precision, and propose strategies to correct for the effect. 
We find that the induced bias becomes a significant fraction of the statistical uncertainty for deviations from the fiducial cosmology, at the level of $|\Delta \Omega_m| = 0.03$ and $|\Delta N_\mathrm{eff}| = 0.3$, and for very precise data corresponding to a forecasted Year-5 DESI survey (or other stage IV dark energy galaxy surveys). We present several ways to correct for this effect, suitable for a variety of applications. We therefore recommend that analyses exploring such parameter regimes either apply the proposed corrections or include an appropriate systematic error budget.
}

\begin{document}
\maketitle

\section{Introduction}
The sound horizon scale ($r_d$) is a fundamental quantity relevant both for the interpretation of cosmic microwave background (CMB) observations and the analysis of baryonic acoustic oscillations (BAO). In the early Universe, the photons and baryons were coupled and behaved as a forced damped harmonic oscillator with gravitational forces attempting to cluster the baryons, the photon pressure keeping structures from collapsing \cite{bao-motivation-10.1093/mnras/stt1687, bassett2009baryon_bao_review, percival2013large}. After the drag epoch, redshift $z_d \sim 1060$, the baryons can start clustering under the effect of gravity, leaving the imprint of these oscillations, whose fundamental mode is set by $r_d$ (the sound horizon at $z_d$) in the dark matter and, in turn, in the large-scale structure (LSS) of the Universe \cite{planck2015, WMAP}. Therefore, although a biased tracer of the dark matter, the distribution of galaxies in the late-time Universe is related to the distribution of baryons in the early Universe (see e.g., \cite[Chapter 5]{Coil_2013}) and carries information about $r_d$. The sound horizon acts as a standard ruler \cite{Eisenstein_2005_BaoStandardRuler+shortcut} both in the CMB (sound horizon at decoupling) and the LSS (sound horizon at radiation drag), allowing us to measure cosmological distances, constrain cosmological parameters, and anchor the inverse distance ladder \cite{BOSS:2014hhw,10.1093/mnras/stv261}. The BAO scale encodes the standard ruler information, manifested as a peak in the galaxy two-point correlation function (hereafter \enquote{BAO peak}) and a series of oscillations in the power spectrum (hereafter \enquote{BAO oscillations}). The way galaxies are distributed both radially (in redshift) and angularly in the sky (right ascension and declination) is accessed through galaxy spectroscopic surveys. Overall, such surveys have been covering increasingly larger cosmological volumes at increasing sensitivity to faint galaxies (see analyses of surveys like the Anglo Australian Two- and Six-Degrees-Field surveys [2dF, \cite{10.1111/j.1365-2966.2005.09318.x}, 6dF \cite{2009MNRAS.399..683J}], the Sloan Digital Sky Survey [SDSS, \cite{Eisenstein_2005_BaoStandardRuler+shortcut}] or the Dark Energy Spectroscopic Instrument survey [DESI, \cite{desicollaboration2024desi2024vicosmological}]). This translated into a greater statistical precision of the inferred cosmological quantities (e.g. 2024 DESI Data Release 2 reaching an aggregate precision of 0.28\% in BAO scale measurement and peak determination \cite{DESI:2025zgx}). With increasing statistical precision, exquisite control of systematic errors becomes paramount, and small effects once negligible, such as universally adopted approximations at the data analysis level, may become important.   

An example of such a systematic uncertainty introduced in the analysis is the computation of the sound horizon $r_d$ itself. The standard analysis of BAO from galaxy clustering data proceeds in two steps: 1) a model-independent measurement of compressed quantities capturing the apparent size of the sound horizon is performed, yielding the BAO shift parameters (usually denoted by $\alpha$), and 2) these BAO shift parameters are translated into cosmological parameters within a given model. While this approach greatly simplifies many aspects of the pipeline, accelerates cosmological parameter inference, and allows for greater interpretability of the results, it also introduces potential avenues for small systematic issues if the analysis is not performed exceedingly carefully. In step 1 above, the extraction of compressed cosmological information from the galaxy survey data (already compressed into summary statistics such as the power spectrum or correlation function) requires fitting a given template model to the data to extract the BAO location while marginalizing over additional information (such as the BAO peak amplitude, the broadband shape, etc.). This compression technique yields the BAO shift parameters (along the line of sight $\parallel$ and across the line of sight $\perp$), 
\begin{equation}
\alpha_\parallel(z) = (D_H(z)/r_d)/(D_H(z)/r_d)^\mathrm{fid}~, \qquad \quad \qquad \quad \alpha_\perp(z) =(D_M(z)/r_d)/(D_M(z)/r_d)^\mathrm{fid}~,
\label{eq:alphas}
\end{equation}

which are normalized cosmological distances (transverse comoving distance $D_M$ and the Hubble distance $D_H$) relative to a fiducial model, indicated by \enquote{fid}. However, the same procedure is not reproduced exactly in the theoretical interpretation of these quantities (step 2). Instead of producing a power spectrum and converting that into a prediction for the scaling quantities, one typically just evaluates the sound horizon in \cref{eq:alphas} through a simple integral \cite{Thepsuriya:2014zda}, $r_d^{\rm int}$\,, and combines it with the computed distances. By doing so, one is indirectly assuming that the decoupling is instantaneous and that no other effects are shifting the BAO oscillations. It has been known as early as 2008 that this is inexact: extracting $r_d$ from the position of the BAO peak in the correlation function is affected by non-linear effects like Silk damping and velocity overshoot \cite{Sanchez:2008iw}. The full effect of this approximation, $r_d\simeq r_d^{\rm int}$\,, on cosmological inference is reduced because inference in practice is always done {\it relative} to a fiducial model (adopting the approximation both in the numerator and denominator of \cref{eq:alphas}). Works studying the dependence of the cosmological inference results on the choice of the fiducial cosmology (or the fiducial template) include \cite{Bernal2020}, who explicitly test the robustness to modifications of the growth of perturbations prior to recombination, and \cite{Perez-Fernandez25}, who evaluate the impact of the fiducial cosmology assumed in the BAO analysis of DESI data. For previous surveys, the effect was found to be small; for DESI, it is propagated into a systematic error budget. 

Here, we focus on a specific component of the analysis that is influenced but not caused by the choice of the fiducial cosmology. Adopting the specific approximation $r_d/r_d^{\rm fid} \simeq r_d^{\rm int}/r_d^{\rm int, fid}$ can, in principle, still introduce small systematic effects ({\it shifts} or {\it deviations} in the recovered cosmological parameters): even in the case where the distances match those of the fiducial model, a shift in $r_d$ induces a shift in $\alpha$, $\Delta r_d/r_d\simeq\Delta \alpha/\alpha$ which affects equally $\alpha_{\parallel}$ and $\alpha{\perp}$. This specific shift has been studied already in great detail in the context of previous surveys, see \cite{Thepsuriya:2014zda, Carter:2019ulk}. Ref.~\cite{Thepsuriya:2014zda} found a 0.15\% deviation for a standard set of parameter variations within $\Lambda$CDM, as well as a slightly larger effect for extra relativistic relics $N_\mathrm{eff}>3$ and massive neutrinos $\sum m_\nu>0$. Instead, \cite{Carter:2019ulk} adopted a simulation-based approach which models the analysis pipeline more completely and studied the impact statistically in the same parameter space, finding an effect up to $0.5\%$. In particular, their study also included the impact of non-linear corrections to the power spectrum (through the use of the Aemulus simulation suite) as well as the impact of BAO density field reconstruction and other pipeline-related effects on the extracted shift quantities. For a detailed comparison to these earlier studies, see \cref{sec:nonlin_compare}.

While these previous studies can be used to estimate the size of the sound horizon mismatch (between that computed through the integral and that actually imprinted in the large scale structure) and to derive a corresponding systematic uncertainty, the parameter space explored does not cover many currently highly investigated models (missing for example those changing the sound horizon through pre-recombination physics) and it is not fully clear which parts of the effects seen in e.g. \cite{Carter:2019ulk} can be mitigated using more modern and advanced modeling approaches. Here, using the same non-stochastic methodology as in \cite{Thepsuriya:2014zda}, we improve on these previous studies by {\it i)} covering an extended parameter space, {\it ii)} specifically investigating the potential systematics in the context of state-of-the-art galaxy redshift surveys such as the DESI survey, and {\it iii)} propose how such effects could be mitigated in future likelihood-based analyses. 

\Cref{sec:methods} focuses on the necessary definitions and a pedagogical introduction to the corresponding background material, while we present the results of our analysis in the context of a survey with the  DESI volume and number density in \cref{sec:results}. We then compare our results to previous studies and use more realistic modeling in \cref{sec:nonlin_compare}. We discuss ways of correcting for this effect in likelihood codes in \cref{sec:correcting}, and we conclude in \cref{sec:conclusions}.

\section{Methods to extract the sound horizon}\label{sec:methods}

We usually identify the theoretical definition of $r_d$ -- an integral over the sound speed in the early Universe -- with the (measured) scale of the BAO signature in galaxy clustering. However, while the theoretical definition depends only on (early universe) background quantities and assumes an instantaneous redshift of radiation drag, the BAO scale as imprinted in the large-scale structure clustering at late-time is affected by a suite of additional effects. Even neglecting non-linearities (non-linear gravitational evolution, bias, non-linear redshift space distortions, velocity bias \cite{Tseliakhovich_Hirata, Sanchez:2008iw}, etc.) and any possible observational effects, the BAO scale in the matter clustering does not coincide with the theoretical prediction of the integral. This section illustrates and quantifies these differences.

This section begins pedagogically with the simplest, idealized, implementations of determining the BAO scale in \cref{ssec:bao_int} and gradually allows complexity to increase until reaching a semi-realistic DESI-like setup in \cref{ssec:desi_bao_analysis_method}.

\subsection{The sound horizon integral}\label{ssec:bao_int}
The sound horizon is usually defined as an integral of the sound speed with respect to conformal time from the Big Bang\footnote{In practice the sensitivity on the precise lower limit is negligible, so we just require an integral to much higher redshift than recombination -- above a redshift of $10^6$ the contribution of the integral to the sound horizon is already only at the permille level.} to the epoch of baryon drag. It can also be written as an integral over the redshift
\begin{equation}\label{eq:rd_int}
    r_d^\mathrm{int} = \int_{z_d}^\infty \frac{c_s(z) \mathrm{d}z}{H(z)}~.
\end{equation}

We have explicitly labeled it $r_d^\mathrm{int}$ in order to differentiate it from the other ways of computing or measuring a sound horizon using the power spectrum (see below). Here $z_d$ is the redshift of baryon drag, $H(z)$ the Hubble parameter, and $c_s(z)$ the baryon-photon sound speed. Note that while the decoupling of the baryons from the photons is a continuous process, the integral formulation, \cref{eq:rd_int}, assumes an instantaneous decoupling.

We can write the sound speed generally as
\begin{equation}\label{eq:soundspeed}
    c_s(z) = \frac{1}{\sqrt{3 \left(1+R(z)\right)}}~,
\end{equation}
where $R(z)$ is the baryon loading, see \cref{eq:R} below. The redshift of baryon drag $z_d$ is defined as the time when the baryon optical depth becomes unity:
\begin{equation}\label{eq:zdrag}
   1 \equiv \int_{0}^{z_d} \frac{\sigma_T n_e(z)}{(1+z) H(z) \, R(z)}\, \mathrm{d}z~.
\end{equation}
with $\sigma_T$ being the Thompson scattering cross section and $n_e$ the electron number density. While these equations are in principle sufficient for computing the sound horizon for a given cosmology, it is not immediately clear which physical effects and therefore which cosmological parameters (for the Universe's constituents) they strongly depend on. This information can be very helpful when quantifying the most important parameter dependencies of the sound horizon.

We can make the dependencies on the cosmological parameters more explicit using the following expressions:
\begin{align}
    R(z) = \frac{3\rho_b(z)}{4\rho_\gamma(z)}  & \approx \frac{670}{1+z} \cdot \frac{(\Omega_b h^2/0.022)}{(T_\mathrm{cmb}/2.7255\mathrm{K})^4}~, \label{eq:R}\\
    \sigma_T n_e(z) & \approx 5.2\cdot x_e(z) (1+z)^3   \cdot \left(\frac{1-Y_\mathrm{He}}{0.75}\right) \cdot \left(\frac{\Omega_b h^2}{0.022}\right)\, [\mathrm{km/s/Mpc}]~,\label{eq:ne}
\end{align}
where $\Omega_bh^2$ denotes the physical density of baryons, $T_{\rm cmb}$ the CMB black body temperature, $Y_{\rm He}$ the (primordial) helium abundance and $x_e(z)$ the free electron fraction at redshift $z$.  \Cref{eq:rd_int,eq:soundspeed,eq:zdrag,eq:R,eq:ne} show explicitly which physics (and cosmological parameters) the sound horizon generally depends on. \Cref{eq:zdrag,eq:ne} together show that the most significant contribution to the integral is during recombination, where $x_e$ sharply increases with redshift. Therefore, the Hubble parameter after the redshift of recombination is \emph{largely unimportant} despite the integral technically starting at $z=0$.\footnote{For the Planck TTTEEE best-fitting cosmology, the redshift of photon last scattering is $1085$ while that of baryon drag is around $1060$, only a $2.3\%$ difference. The disambiguation between recombination and the baryon drag epoch is therefore not crucial to this specific discussion.} Besides this, the sound horizon integral in \cref{eq:rd_int} depends only on quantities at $z > z_d$, therefore completely eliminating the dependency on any late-universe contributions, such as for example curvature ($\Omega_k$) or dark energy ($\Omega_\mathrm{DE}(z)$). It also does not depend on the distribution of the primordial perturbations (albeit we focus on adiabatic initial conditions), and therefore neither $A_s$ nor $n_s$ (or running, or features in the primordial power spectrum) have any impact on the sound horizon.

In the absence of very non-trivial physics impacting the photons and baryons, the sound speed is entirely characterized by $T_\mathrm{cmb}$ and $\Omega_b h^2$. Similarly, the drag redshift is also determined by these two quantities at leading order (mostly through $x_e$), except if there is a delay/advancement of recombination (see \cref{ssec:me}). The Hubble parameter before the drag redshift in most models is dominated by $\Omega_m h^2$ and $\Omega_r h^2 = \Omega_\gamma h^2 (1+7/8 \cdot N_\mathrm{eff} \cdot (4/11)^{4/3})$, and possibly other contributions to the energy density beyond $\Lambda$CDM, such as early dark energy (see \cref{ssec:ede}). Here $N_\mathrm{eff}$ is the usual effective number of relativistic species, which in $\Lambda$CDM is constant and equal to 3.044 \cite{Drewes:2024wbw}.

In summary, the dominant parameter dependencies of the sound horizon are
\begin{equation}
    \{\Omega_b h^2, \Omega_m h^2, T_\mathrm{cmb}, N_\mathrm{eff}\}~,
\end{equation}
as well as modifications in early universe physics; either for shifts in recombination redshift or when changing the Hubble parameter through additional contributions in the pre-recombination universe.

\begin{figure}
    \centering
    \includegraphics[width=0.64\linewidth]{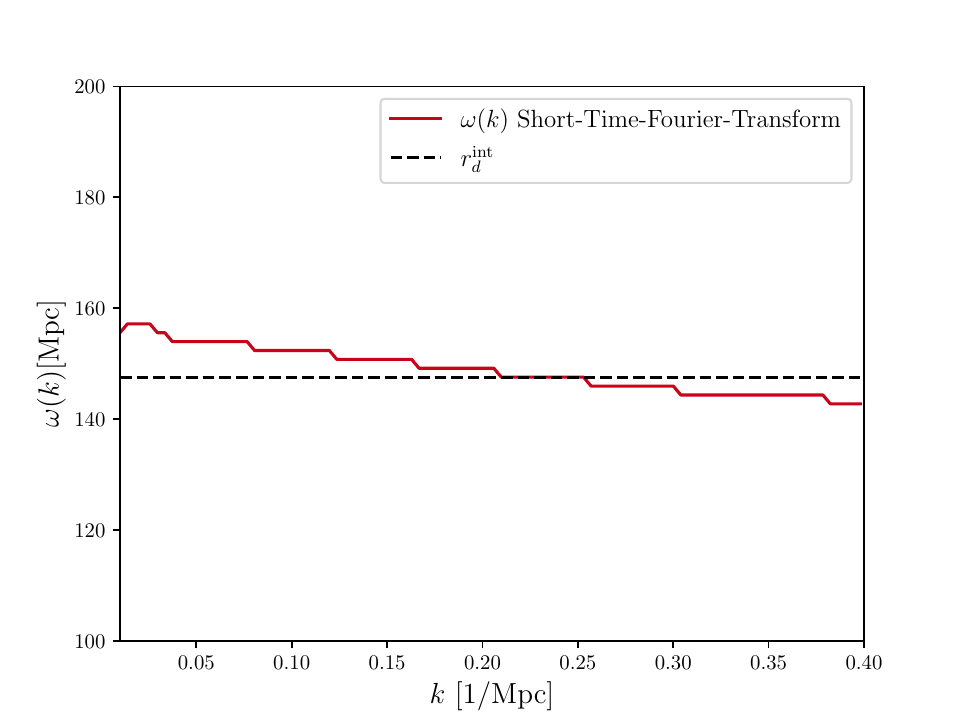}
    \caption{Analysis of the BAO frequency as a function of the scale for a cosmology with $\Omega_m=0.3$, $\Omega_b=0.05$, and $h=0.68$ ($r_d^\mathrm{int} = 147.46$Mpc). We compute the linear power spectrum using \texttt{class} \cite{class} and recover the oscillations using the de-wiggling method presented in \cite{Ghaemi:2025lgu} (\enquote{Cubic Inflections}). Then, we use a Short-Time-Fourier-Transform from the \texttt{tftb} package (\url{https://github.com/scikit-signal/tftb}) to compute the instantaneous frequencies. We also show a comparison to the expected oscillation frequency $\omega=r_d^{\rm int}$ from \cref{eq:rd_int}. Modified version of \cite[Fig.~2.4]{Schoneberg:2021uak}. The visible steps in the graph represent the bin/window size of the Short-Time-Fourier-Transform.}
    \label{fig:rd_int_vs_stft}
\end{figure}

\subsection{The BAO in the linear power spectrum}\label{ssec:bao_power}

The sound horizon $r_d^{\rm int}$, plays a crucial role in describing the analytical approximation of the BAO, which can be approximated for the Fourier power spectrum to first order as $\sin(k r_d^\mathrm{int})$, see for example \cite{Hu:1995en,Eisenstein:1997ik}. Therefore, it should in principle be possible to extract the oscillation frequency directly from the (damped) BAO wiggles in the power spectrum. However, it has been shown for example in \cite{Montanari:2011nz} (based on \cite{Hu:1995en,Eisenstein:1997ik}) that deviations of the peak positions compared to the simplest expectation given by $\sin(k r_d^\mathrm{int})$ are expected in the full analytical calculations -- hence the oscillation frequency $\omega$ is not $r_d^{\rm int}$.  These deviations differ for the various peaks and effectively result in a scale-dependent oscillation frequency, see for example \cite[fig.~2.4]{Schoneberg:2021uak}, which we reproduce for convenience in slightly modified form in \cref{fig:rd_int_vs_stft}. It is evident that the oscillation frequency is not constant and changes with wavenumber. This is a first indication that the standard ruler of the BAO does not appear as a perfectly standard feature in the galaxy power spectrum. Different approaches to measuring the BAO standard-ruler signal from the large-scale structure power spectrum may be affected by this in different ways.

In the following, we build up a set of three ways of extracting the sound horizon from the power spectrum, of increasing complexity and increasingly closer to the methodology used in actual analysis pipelines of galaxy surveys.

An important tool for this discussion is the decomposition of the power spectrum into an oscillatory part and a broadband part. We schematically write $P(k) = P^\mathrm{no-wiggle}(k) + P^\mathrm{BAO}(k)$. This decomposition is typically performed using a de-wiggling algorithm, see \cite{Ghaemi:2025lgu} for a comparison and discussion of different algorithms.

\paragraph{Peak positions (peak):} Instead of fitting the BAO oscillations of the power spectrum, we start by focusing on the peak (BAO oscillation maxima) positions to estimate the sound horizon. 
As we will see below, this oversimplified approach does not provide sufficiently accurate results, and we use it only for illustrative purposes.

After subtracting the de-wiggled power spectrum $P^\mathrm{no-wiggle}(k)$ from a given power spectrum using the de-wiggling method Cubic Inflections presented in \cite{Ghaemi:2025lgu}, we fit the linear relation expected from $\sin(k \omega)$ through the peaks of the resulting oscillations. In this case, the slope of the relation between $\pi (n+\frac{1}{2})$ and $k_n$ (where $k_n$ is the $n$-th peak/valley of the BAO) determines the value of the extracted sound horizon, here denoted as $r^\mathrm{peak}_d$. Note that the linear relation is fit \emph{without} any phase offset.\footnote{This is motivated by the analytical approximation of $\sin(k r_d)$ \cite{Eisenstein:1997ik}, with peaks/troughs at $k_n = \pi (n+\frac{1}{2})/r_d$, which can be used to determine the slope as $r_d = \pi (n+\frac{1}{2})/k_n$\,. Given that the peak at $n=0$ coincides with the peak of the power spectrum (which is, for most BAO de-wiggling methods, not included, or inaccurately determined, see \cite{Ghaemi:2025lgu}), we do not include it. Finally, in a version with offset, one would allow for a small phase shift $\sin(k r_d+\phi)$, but we found numerically that this approach is even less accurate. We also put a threshold on the size of the peaks at $20\%$ of the largest peak in order to avoid fitting numerical artifacts of the de-wiggling procedure or peaks that are too damped to be well-determined in position.} 

Testing this method on a cosmology with $\Omega_m=0.3$, $H_0 = 68$km/s/Mpc, $\Omega_b = 0.05$, the resulting $r_d^\mathrm{peak} = 153.42\, \mathrm{Mpc}$ is very different ($-3.8\%$) from the $r_d^\mathrm{int} = 147.46\, \mathrm{Mpc}$, due to the drift in oscillation frequency (see \cref{fig:rd_int_vs_stft}). However, the relevant quantity for cosmological inference is the ratio of the sound horizon to the fiducial, $s^{\rm peak} = r_{d, \mathrm{fid}}^\mathrm{peak}/r_d^\mathrm{peak}$. In this case, the two relative sound horizons $s^{\rm peak}$ and $s^{\rm int}$ are in much better agreement, see \cref{sec:results}. This is an important consideration that is at the basis of all current BAO analyses: while different methods might return significantly different $r_d$ values, the corresponding relative variation in response to changes in cosmological parameters will be very similar, and therefore the values of $s$ will be much more consistent.

\paragraph{Only oscillations (BAO):}
There are several further issues with the previous method. Assigning uncertainties correctly is very difficult (as is common to all methods that rely on isolating specific locations or points in the data space). Additionally, the wavenumber dependence of the oscillation frequency means that trying to extract a single BAO frequency is ill-defined \textit{a priori}. As such, we need a method that operates directly on the power spectrum over a range of wavenumbers $k$ and which extracts primarily how the wavenumber-dependent oscillation frequencies of the observed and fiducial cosmology \textit{differ}. We focus here on using only the BAO oscillations, extracted as discussed above.

We create a template of the BAO for a fiducial cosmology by subtracting the de-wiggled power spectrum (using the same technique as above) from the full power spectrum. Then, the (fiducial) template is shifted according to the parameter $s$, which is adjusted to fit the BAO of the cosmology under consideration. To incorporate the measurement uncertainties in this process, we use a Gaussian likelihood $\mathcal{L}$, for which $\chi^2 = - 2 \ln \mathcal{L}$ can be written as
\begin{equation}\label{eq:likelihood}
    \chi^2(s) = \frac{V}{4\pi^2} \int \frac{\left[P(k)-P^\mathrm{fid}(k s)\right]^2}{\left[\mathcal{N}+P^\mathrm{fid}(k)\right]^2}  k^2 \mathrm{d}k\approx \frac{V}{4\pi^2}\int \frac{\left[P^{\rm BAO}(k)-P^\mathrm{BAO, \text{fid}}(k s)\right]^2}{\left[\mathcal{N}+P^\mathrm{fid}(k)\right]^2}  k^2 \mathrm{d}k~,
\end{equation}
where $P(k)$ is the power spectrum of the considered cosmology, $P^{\rm fid}$ is the fiducial power spectrum,
$s$ is the corresponding shift, $V$ is the survey volume, and we approximated the covariance through a stochastic noise term $\mathcal{N}$ in addition to the fiducial power spectrum $P^\mathrm{fid}(k)$. 

The $\approx$ sign marks the (incorrect) assumption that the broadband cancels out in the numerator. In practice, this requires marginalization over a nuisance parameter rescaling the amplitude of the oscillations (see also \cref{ssec:null}) in order to make this method well-behaved.

The range used to perform the fit of the power spectrum is $k \in [0.05,0.25]$ Mpc$^{-1}h$ for the fiducial model, which includes most of the BAO features. For any other model, we scale this range by the ratio $r_d^\mathrm{int}/r_d^\mathrm{int,fid}$, which is the leading order expectation of the shift of the BAO. We estimate the recovered $s$ as that which minimizes the $\chi^2$, and its uncertainty by imposing $\Delta \chi^2(s) = \chi^2(s) - \chi^2_\mathrm{min}= 1$  following the Frequentist idea of the likelihood ratio based on Wilk's theorem (the \enquote*{graphical method} in \cite{Herold:2024enb}). Typically, we expect that uncertainties constructed in this way do not deviate strongly from their Bayesian counterparts, given the simple Gaussian form assumed for the likelihood in \cref{eq:likelihood}.

We do not incorporate many of the systematic corrections that real data require, and we do not consider redshift space distortions (and therefore we do not need to model the 
Kaiser effect, Fingers of God, etc., or to project into the Legendre multipoles). 
While this is certainly a simplified approximation of the modeling used in a full data analysis pipeline, it captures the relevant aspects and is sufficient for quantifying the difference between $r_d^{\rm int}$ and the observed $r_d$, as portrayed in \cref{sec:results,sec:nonlin_compare}. We denote the $r_d$ estimate from this method as $r^{P(k)~\rm BAO}_d$.

\paragraph{Full modeling:} The previous approach relies on the approximate removal of the broadband in the linear power spectrum. As we will see in \cref{sec:results}, this \enquote{de-wiggling} can potentially introduce a non-negligible bias in the final estimate of the sound horizon. A preferable method would be insensitive to broadband information without relying on de-wiggling. Hence, here we use the full linear power spectra, and a Gaussian likelihood given by
\begin{equation}\label{eq:likelihood_full}
    \chi^2(s) = \frac{V}{4\pi^2} \int \frac{\left[P(k)-P^\mathrm{template}(k s)\right]^2}{\left[\mathcal{N}+P^\mathrm{fid}(k)\right]^2}  k^2 \mathrm{d}k~,
\end{equation}
with the same range of $k \in [0.05,0.25]$ Mpc$^{-1}h$ rescaled by the sound horizon integral ratio as above. The template power spectrum is modeled as\footnote{Since the coefficients $a_n$ can be rescaled, it does not matter if in \cref{eq:pk_template} we write $P^\mathrm{template}(rk)$ or $P^\mathrm{template}(k s)$.}
\begin{equation}\label{eq:pk_template}
    P^\mathrm{template}(k) = B \cdot P^\mathrm{fid}(k) + a_1 k^2 + a_2 k + a_3 + a_4/k + a_5/k^2~,
\end{equation}
where the parameters $B$ and $a_n$ are additional nuisance parameters to remove broadband information, which would otherwise contain additional cosmological information beyond that extracted in traditional BAO analyses. These nuisance parameters would be marginalized over in a Bayesian analysis, or in this case, for a Frequentist estimate, they are set to the values that minimize the $\chi^2(s)$. For each value of $s$ we minimize the $\chi^2(s)$ over all of the nuisance parameters, and use the same $\Delta \chi^2(s)=1$ criterion as before to estimate errors. While again this method is not a faithful reproduction of the full data pipeline, it captures the main aspects of the physical signal, as we confirm in \cref{sec:nonlin_compare}. It is thus sufficient for quantifying the effects of differences between $r_d^{\rm int}$ and $r_d^{\rm obs}$, which for this implementation we refer to as $r^{P(k)~\rm full}_d$\,. More advanced modeling, such as the one employed in \cite{Carter:2019ulk}, will typically introduce \textit{additional} deviations in the extracted $\alpha$ parameter(s) but are unlikely to reduce the effects found here (see \cref{sec:nonlin_compare}).

\subsection{The BAO in the correlation function, \texorpdfstring{$\xi$}{xi}}\label{ssec:bao_cf}

\begin{figure}
    \centering
    \includegraphics[width=0.7\linewidth]{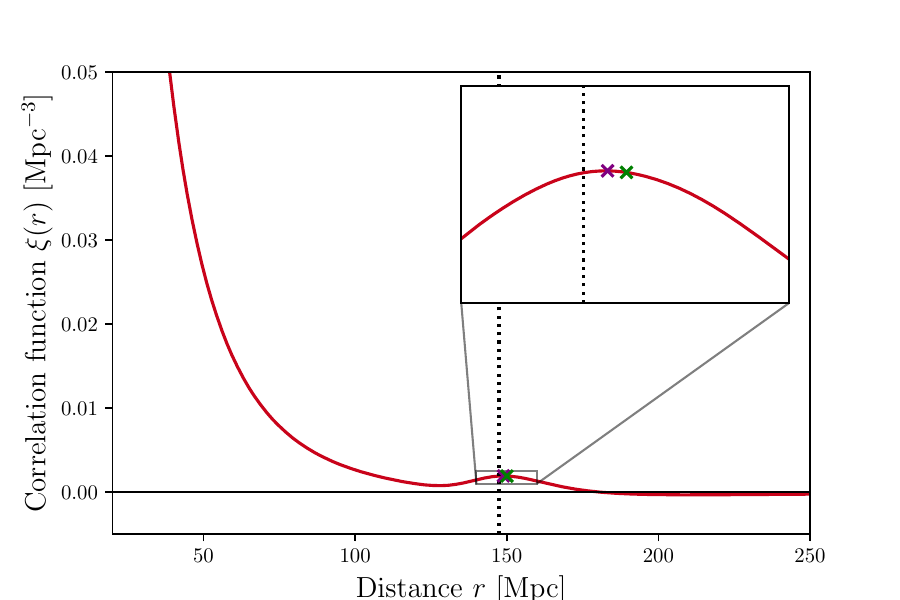}
    \caption{Correlation function for the same cosmology as in \cref{fig:rd_int_vs_stft}. We show in a black dashed line the value of $r_d^\mathrm{int} = 147.46$ Mpc, while the purple marker shows the peak of the correlation function, and the green marker shows the peak of the correlation function once the broadband has been subtracted. The peak of the correlation function is at $148.93$ Mpc (+1\%) without broadband subtraction and at $150.08$ Mpc (+1.8\%) with broadband subtraction (using the same polynomial broadband of \cref{eq:cf_template}).}
    \label{fig:rd_int_vs_CF}
\end{figure}
Because in the correlation function, the BAO signature is localized in a single peak, one may expect analyses using the correlation function to be comparatively simpler. However, the exact position of the peak of the correlation function is shifted from the naive expectation by several effects (such as, for example, the linear-theory Silk damping or the non-linear evolution, see \cite{Sanchez:2008iw}). We show such a shift in \cref{fig:rd_int_vs_CF} for the linear case, where the markers represent the peak of the correlation function (with and without the broadband removed), and the dotted line represents the value of $r_d^\mathrm{int}$. 

Therefore, as for the power spectrum, the analysis employs a template which is shifted according to the parameter $s$ and the broadband differences are absorbed by polynomial term corrections.

Inspired by the method developed in \cite{Thepsuriya:2014zda} we write\footnote{Since the coefficients $a_n$ can be rescaled, it does not matter if in \cref{eq:cf_template} we write $\xi^\mathrm{template}(r)$ or $\xi^\mathrm{template}(r/s)$.}
\begin{equation}\label{eq:cf_template}
    \xi^\mathrm{template}(r) = B \cdot \xi^\mathrm{fid}(r) + a_1 + a_2 /r + a_3 /r^2 + a_4/r^3 + a_5 /r^4 +a_6/r^5~,
\end{equation}
where the $B$ and $a_n$ are nuisance parameters which are being marginalized/minimized over. The (log) likelihood in this case is given by 
\begin{equation}
    \chi^2(s) = [\xi(r)-\xi^\mathrm{template}(r/s)]^T C^{-1} [\xi(r) - \xi^\mathrm{template}(r/s)]~,
\end{equation}
with the correlation function $\xi(r)$ for the given cosmology, the template $\xi^\mathrm{template}(r)$ shifted by the shift parameter $s$, and the covariance matrix $C$ determined as in \cite[Eq.~(9)]{CF_cov}. The range used to fit the fiducial model spans $r \in [100,200]$ Mpc, since the peak tends to be located in this range. For other models, we rescale the fiducial range by the (inverse of the) ratio of $r_d^\mathrm{int}/r_d^\mathrm{int, fid}$ as before. Similarly to the power spectrum case, we recognize that this is a simplified approximation, but it is sufficient to forecast the effects of the main differences between $r_d^{\rm int}$ and $r_d^{\rm obs}$, which in this implementation we refer to as $r_d^{\xi(r)}$\,.

\subsection{DESI-like BAO power spectrum analysis} \label{ssec:desi_bao_analysis_method}
In some cases, we consider an even more realistic scenario, incorporating also the impact of non-linear evolution on the power spectrum through the use of effective field theory (EFT) as well as certain redshift-space related effects. In this way, we can determine if the effects we are studying persist in a more realistic analysis or are absorbed by one of the more advanced nuisance terms that are part of the state-of-the-art pipelines. To test this, we have built a pipeline similar to the one used to obtain the official DESI results, following the recommendations of \cite{DESI2024III_Adame_2025}. 

We generate a (mock) data vector containing the monopole, quadrupole, and hexadecapole ($\ell = 0,2,4$) of the nonlinear galaxy power spectrum obtained with the EFT code \texttt{velocileptors} \cite{velocileptors_i_Chen_2021, velocileptors_ii_Chen_2020}. We evaluate the multipoles in $k\in [0.02, 0.3]$ h Mpc$^{-1}$ with 28 equidistant points sampled. We don't use BAO reconstruction at any point during the process (unlike \cite{Carter_2020} or the official DESI results) in order to keep the model relatively simple and to aid with the analytical understanding. The covariance used in this analysis is analogous to the one used above (Gaussian, cubic volume).

The model used to \emph{fit} the mock data vector consists of a BAO template defined as (c.f. \cite[e.q.~(4.4)]{DESI2024III_Adame_2025})
\begin{equation}
P_{\ell, \text{obs}}(k) = \frac{2\ell + 1}{2} \int_{-1}^{1} d\mu \Bigg[\, \mathcal{L}_{\ell}(\mu) \big[ \mathcal{B}(k, \mu) P^\mathrm{no-wiggle}(k) + \mathcal{C}(k', \mu') P^\mathrm{BAO}(k', \mu') \big]\Bigg] + \mathcal{D}_{\ell}(k)~.
\label{eq:BAO_template}
\end{equation}

Here, $\ell$ stands for the order of the multipole considered (i.e. $P_{\ell=0}$ is the monopole), $\mu$ is the cosine of the observer's line-of-sight angle, and $\mathcal{L}_{\ell}(x)$ stands for the Legendre polynomial of order $\ell$. $P^\mathrm{no-wiggle}(k)$ and $P^\mathrm{BAO}(k, \mu)$ are computed from the linear power spectrum using the de-wiggling method presented in \cite{wallisch2018}. The Alcock-Pacynski (AP) effect is implemented as usual using

\begin{equation}
     k' = k \cdot \frac{\mu}{\mu' q_\parallel}~, \qquad
     \mu' = \frac{\mu}{\left[\left(\frac{q_{\parallel}}{q_{\perp}}\right)^2 (1-\mu^2)+\mu^2\right]^{1/2}}~,
\end{equation}
with the magnitudes $q_\parallel$ and $q_\perp$ defined by
\begin{equation}
    q_{\parallel}(z) = \frac{D_H(z)}{D_H^{\rm fid}(z)}~, \qquad q_{\perp}(z) = \frac{D_M(z)}{D_M^{\rm fid}(z)}~,
\end{equation}
analogous to the shift parameters $\alpha_\parallel$ and $\alpha_\perp$, just without the sound horizon ratio. As evident in \cref{eq:BAO_template}, the AP scaling is only applied to the BAO term as the broadband information is marginalized over.

For the rest of the terms introduced in \cref{eq:BAO_template} (i.e. $\mathcal{B}(k, \mu)$, $\mathcal{C}(k, \mu)$, and $\mathcal{D}_\ell(k)$) note that the first two terms are the same for all the multipoles considered, while the third one is different for each multipole. We follow the definitions proposed by \cite{DESI2024III_Adame_2025}:
\begin{itemize}
    \item The broadband part of the power spectrum considered is modulated by
    \begin{equation}
        \mathcal{B}(k, \mu) = (b_1 + f\mu^2)^2\left(1+\frac{1}{2}k^2\mu^2\Sigma_s^2\right)^{-2}
    \end{equation}
    where the first term is a generalisation of the Kaiser factor, and in the second term, we essentially take into account the effect of Fingers of God (FoG) with a smoothing factor $\Sigma_s$\,; $f$ here represents the logarithmic growth rate of LSS and $b_1$ the linear bias.

    \item The BAO wiggle is anisotropically modulated by 
    \begin{equation}
        \mathcal{C}(k, \mu) = (b_1 + f\mu^2)^2 \exp\left[ -\frac{1}{2}k^2\left(\mu^2\Sigma_{\parallel} + \left(1-\mu^2\right)\Sigma_{\perp}\right)\right]~,
    \end{equation}
    where $\Sigma_{\perp}$ and $\Sigma_{\parallel}$ modulate the damping term for the modes across and along the line-of-sight, respectively.
    \item The last term, $\mathcal{D}_\ell$\,, accounts for both the change in broadband and any residual terms (including e.g., potential counter-terms of the EFT). 
    It is defined in the baseline analysis as a piecewise cubic spline (PCS) like
    \begin{equation}
        \mathcal{D}_\ell = \sum_{n=-1}^{n^{\prime}}a_{\ell, n}W_{3}\left(\frac{k}{\Delta}-n\right)
    \end{equation}
    In this definition, $a_{\ell, n}$ is a free parameter, $W_3(x)$ is the PCS kernel used for the fit (see \cite{pcs_kernel_w3_10.1093/mnras/stw1229}), and $\Delta$ is a term that regulates the \textit{finesse} of the kernel.\footnote{This $\Delta$ parameter is of paramount importance, since it needs to be set to a value that does not allow $\mathcal{D}_\ell$ to replicate the BAO signature. For a discussion on the issue, see \cite{DESI2024III_Adame_2025}.} Using a PCS for $\mathcal{D}_\ell$ instead of the more frequently used polynomial basis is motivated by the results of \cite{DESI2024III_Adame_2025}, and also on several tests carried out by us, where the residuals were greater when taking the polynomial-based approach.
\end{itemize}

The likelihood that we build with this model and the covariance is fundamentally identical to that of \cref{eq:likelihood_full}.

All the steps described, from the data generation to the likelihood definition, are built with \texttt{desilike}\footnote{\url{github.com/cosmodesi/desilike}} and \texttt{cosmoprimo}\footnote{\url{github.com/cosmodesi/cosmoprimo}}. We analytically marginalize the broadband terms of the BAO template likelihood (i.e. all the terms $a_{\ell, n}$) for a faster sampling of the likelihood. The minimization to obtain the maximum a posteriori (MAP) is performed with \texttt{iminuit} \cite{iminuit}, while the Monte Carlo Markov chains (MCMC) are run using \texttt{cobaya} \cite{cobaya_ascl_2019ascl.soft10019T, cobaya_paper_Torrado_2021}. The minimization is run for 30 iterations to ensure convergence of results. The MCMC chain is run until $|\hat{R}-1|<0.01$, where $\hat{R}$ is the Gelman-Rubin convergence criterion \cite{gelman1992inference}, following DESI's standards. We denote the sound horizon obtained in this way as $r_d^\mathrm{DESI}$.

\section{Impact for cosmology inference from current galaxy redshift surveys}\label{sec:results}

In this section, we compare the various ways of computing the sound horizon introduced in \cref{sec:methods} and quantify the systematic shifts in cosmological inference. Our baseline for this comparison is the computation through the integral, denoted by $r^{\rm int}_d$, which is what is currently employed in most pipelines when converting constraints on the compressed parameters $\alpha_\parallel$ and $\alpha_\perp$ into constraints on cosmology. The four alternative ways of obtaining a sound horizon from the data investigated within this section are $r_d^{\rm peak}$, $r_d^{\rm P(k) BAO}$, $r_d^{\rm P(k) full}$, and $r_d^{\xi}$ (collectively referred to as $r_d^{\rm obs})$. Note that we focus on the linear power spectra and the linear sound horizon estimators for this section, leaving the more advanced non-linear case to \cref{sec:nonlin_compare}. As we will see in \cref{sec:nonlin_compare}, these linear cases capture the main effects at the power spectrum level (without reconstruction). The inference of cosmology through $\alpha_\parallel\,$ and $\alpha_\perp$ only depends on the ratios $s^{m}=r_d^{m,\mathrm{fid}}/r^{m}_d$ (for a given method $m$ running through \enquote{int}, and the 4 different type of \enquote{obs}) which are typically more consistent, robust, and stable compared to the individual definitions of $r^m_d$\,, as discussed in \cref{sec:methods}. For $\alpha$ and $s$, we use the same superscripts $m$ -- int, peak, $P(k)$~BAO, $P(k)$~full, $\xi(r)$, and DESI -- as for $r_d$ to indicate the adopted methodology. In what follows, for our simplified approach  we define $\Delta s/s=(s^m-s^{\rm int})/s^{\rm int}$ and so the bias in cosmological inference is given by $\Delta s/s=\Delta \alpha_{\parallel}/\alpha_{\parallel}=\Delta \alpha_{\perp}/\alpha_{\perp}\equiv \Delta \alpha/\alpha$. We determine the statistical uncertainty of $\Delta s/s$ and $\Delta \alpha/\alpha$ for a given survey as described in \cref{sec:methods}.

To minimize biases in cosmological inference, the response of $s^{\rm int}$ to a change in cosmological parameters with respect to the fiducial should be as close as possible to the corresponding response of $s^{\rm obs}$ as extracted from the power spectrum or correlation function. For small enough changes around the fiducial model, such bias can be kept well below the statistical uncertainties and thus be negligible. The important question is: how far away from the fiducial model can parameter exploration go before the biases become a concern. One somewhat arbitrary but widely used \cite{desiv} measure is when the bias reaches 1/5 of the statistical uncertainty. That is when the change in a given cosmological parameter, $\Delta x$, induces a (systematic) shift in the recovered BAO shift parameters $\Delta \alpha$ equal to 1/5 of their statistical error: $\Delta \alpha=\frac{1}{5}\sigma_\alpha$.

As we will show below, while the dependence of $s$ or $\alpha$ on the cosmological parameter of interest can be non-trivial, the \emph{bias between} the integral and a given method $m$ can be well approximated by a linear dependence on a given cosmological parameter, and we discuss where higher order corrections are required in \cref{sec:correcting,app:correction_accuracy}. Therefore, in what follows, we fit a linear relation to the obtained bias $b\equiv \Delta \alpha/\alpha = s^m/s^\mathrm{int}-1$ in response to a change of a given cosmological parameter $x$ and obtain the slope of this relation, $\partial b/\partial x \simeq \Delta b/\Delta x$ which we refer to as the \enquote{Slope of the Bias} (c.f. \cref{tab:bias_analysis}).

We first vary parameters of the $\Lambda$CDM model and then move to those $\Lambda$CDM extensions that modify the sound horizon -- note that these include pre-recombination models; most post-recombination models have little to no impact on the sound horizon. The surveys we investigate are listed in \cref{tab:surveys} (volumes and shot noises, used to determine the statistical errors; HUGE is only used in appendix \ref{sec:forecast}. To allow for a direct comparison with \cite{Thepsuriya:2014zda}, we adopt the same fiducial cosmology they use, consisting of a flat $\Lambda$CDM model with the following parameters:
\begin{equation}\label{eq:fiducial}
    \{\Omega_\mathrm{cdm} h^2 = 0.1196 , \Omega_b h^2 = 0.02207, H_0 = 67.4 \mathrm{km/s/Mpc}, N_\mathrm{eff} = 3.046\}~,
\end{equation} 
and massless neutrinos. This cosmology has $r_d^\mathrm{int} = 147.56$Mpc (see also \cite{Thepsuriya:2014zda}). Note that current theory calculations from BBN yield $N_\mathrm{eff}$ in $\Lambda$CDM closer to 3.044 \cite{Drewes:2024wbw}, but this slight change has only a relative impact of order $6\cdot 10^{-6}$ on the sound horizon, which we consider negligible.

While we performed the analysis for all of the volumes shown in \cref{tab:surveys}, we only show in the following sections the results obtained for the Y5 Total sample (and, in some cases, the Y1 Total sample). See \cref{app:further_results} for these further results. The constraining power is typically dominated by the LRG sample.

\begin{table}[t]
    \centering
    \begin{tabular}{|c|c c c|}
    \hline
        Survey name & Volume [$\mathrm{Gpc}^3h^{-3}$] & Shot noise [$\mathrm{Mpc}^3h^{-3}$]& $P_0(k=0.14/\mathrm{Mpc}) [\mathrm{Mpc}^3h^{-3}]$\\ \hline\hline
        SDSS III & 4.45 & 4400 & 20000 \\ \hline
        DESI Y1 LRG & 3.48 & 5000 & 9000 \\
        DESI Y1 ELG & 1.41 & 6000 & 3000 \\
        DESI Y1 QSO & 0.45 & 40000 & 5000 \\ \hline
        DESI Y5 LRG & 20 & 2700 & 9000 \\
        DESI Y5 ELG & 7.5 & 5000 & 3000 \\  
        DESI Y5 QSO & 2.5 & 40000 & 5000 \\ \hline
        DESI Y1 Total & 5.4 & 5500 & 9000 \\
        DESI Y3 Total & 18 & 4000 & 9000 \\
        DESI Y5 Total & 30 & 3100 & 9000 \\ \hline
        HUGE & 1000 & 3000 & 10000 \\
        \hline
    \end{tabular}
    \caption{Survey specifications used in this work. The specifications are not too dissimilar from those of the named survey in the first column. We note that these deviate slightly from \cite{DESI:2024uvr}. For the DESI~Y3 and Y5 values, we multiply the volumes by $\sim3$ and$\sim5$, respectively, while keeping $P_0$ constant. The shot noise values for the \textit{total} samples have been obtained as described in \cref{app:shot_noise}. The $P_0$ considered for the combined surveys has been approximated as the $P_0$ of the largest sample for each year (LRGs) because of the weight it has on the combined survey.}
    \label{tab:surveys}
\end{table}

\subsection{Null test: Varying \texorpdfstring{$A_s$}{As}}\label{ssec:null}
\begin{figure}
    \centering
    \includegraphics[width=0.49\linewidth]{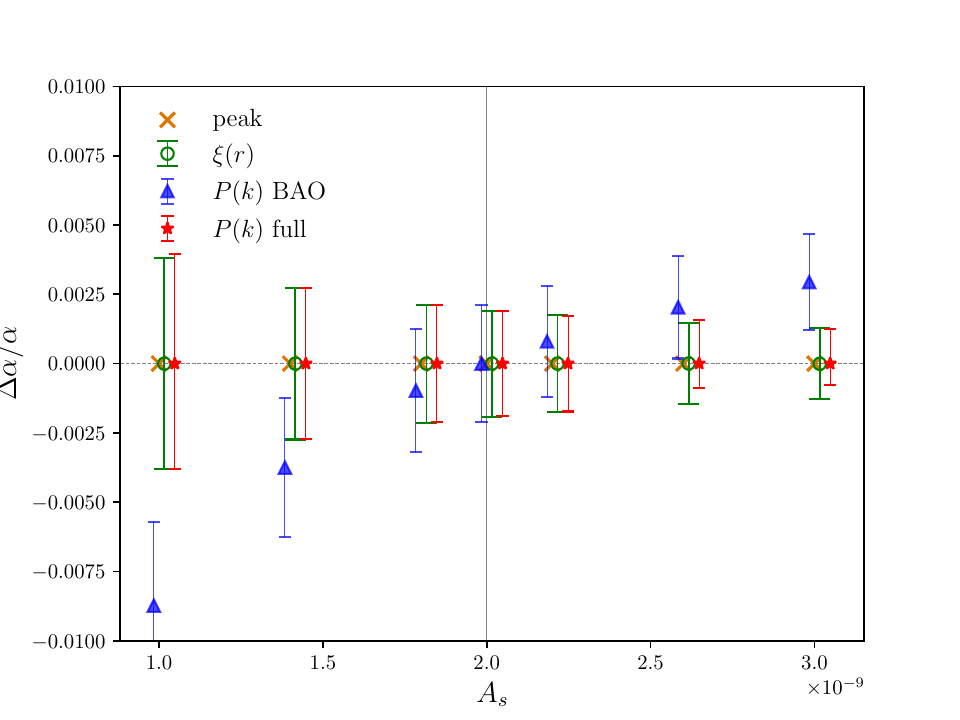}
    \includegraphics[width=0.49\linewidth]{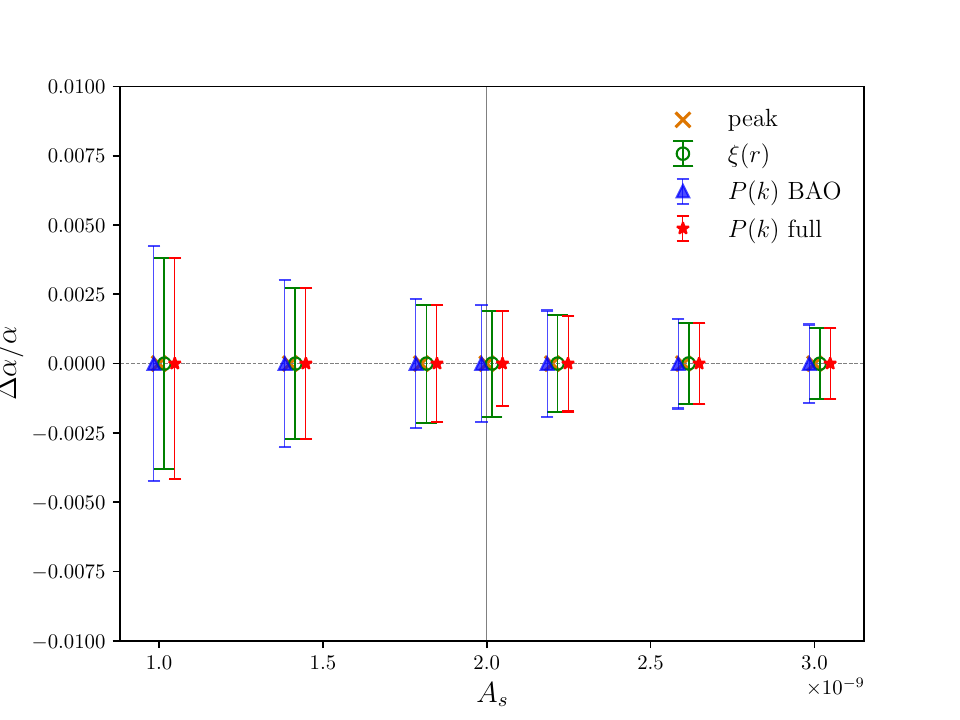}
    \caption{Relative bias on $\alpha$ for the various methods of \cref{sec:methods}
    as a function of the amplitude $A_s$\, with the vertical line indicating the fiducial value; the errors are forecasts for the DESI Y5 (total) survey specification.
    Left: Without the BAO amplitude as a nuisance parameter. Right: with a nuisance parameter for the BAO only method. Points are slightly offset horizontally for clarity.}
    \label{fig:A_s}
\end{figure}

We expect variations of the primordial amplitude $A_s$ to have no impact at all (see also \cref{sec:methods}). We choose a range of $A_s$ significantly larger than recent experiments allow (e.g. \cite{Planck:2018vyg}) to validate this. This is indeed what we observe in the left panel of \cref{fig:A_s} for most methods, except for direct extraction from the BAO oscillations which 
can be biased if the BAO amplitude is not rescaled. This is what motivated us to introduce the BAO amplitude as a nuisance parameter to be marginalized/minimized.
With this (right panel of \cref{fig:A_s}) we find indeed no discernible effect of varying $A_s$\,, except (not unexpectedly) for the size of the error-bars.

In the rest of the manuscript for the \enquote{$P(k)$~BAO} case, we always marginalize over the BAO amplitude. We show in \cref{fig:A_s} only the DESI~Y5 (total) forecast, though we mention that all other surveys of \cref{tab:surveys} show the same (or better) level of consistency.

\subsection{Varying \texorpdfstring{$\Omega_b h^2$}{Omegabh2}}
\begin{figure}
    \centering
    \includegraphics[width=0.48\linewidth]{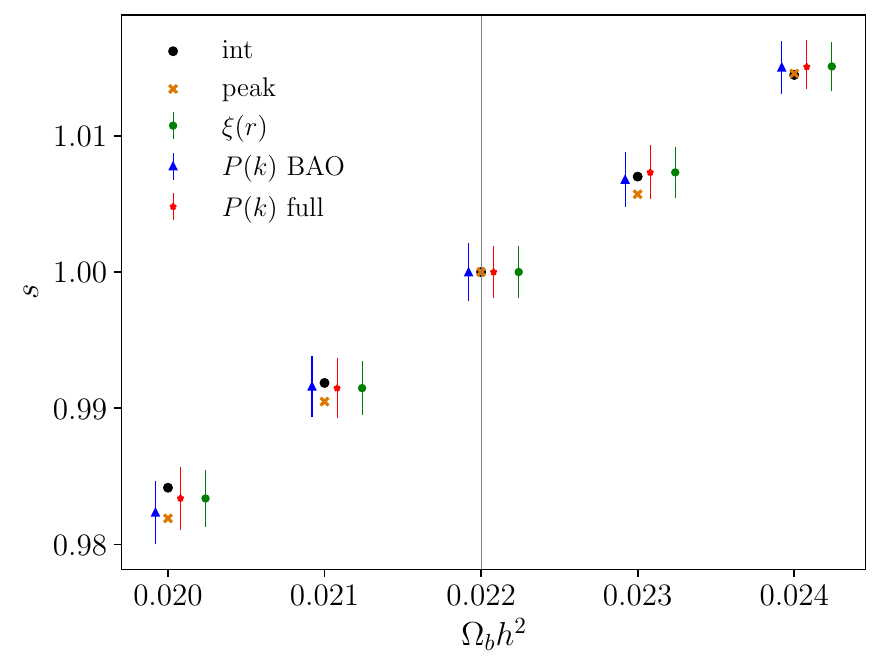}
    \includegraphics[width=0.48\linewidth]{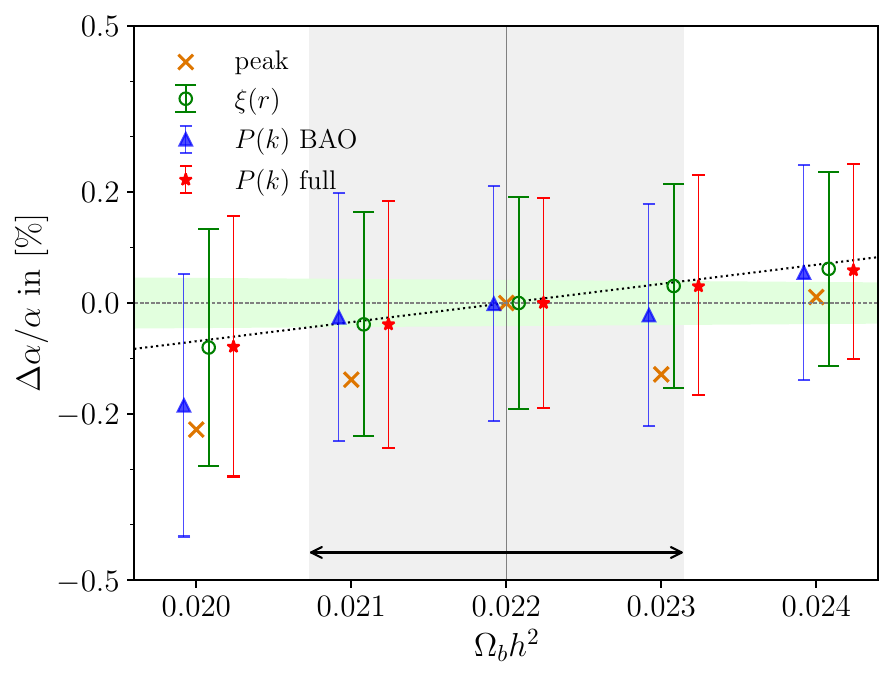}
    \caption{Dependence of the various ways of obtaining $s$ of \cref{sec:methods} on $\Omega_b h^2$ (left) and resulting $\Delta \alpha/\alpha$ relative to the one obtained from the integral definition of $r_d$ (right). Points are slightly offset horizontally for clarity. On the right panel, the horizontal green bands denote the region where the systematic shift in $\alpha$ is below 1/5 of the statistical error. The gray vertical band encloses the range of the cosmological parameter that satisfies this condition.}
    \label{fig:omegab}
\end{figure}

The next simplest case is the variation of $\Omega_b h^2$ (while fixing $\Omega_\mathrm{cdm} h^2$). This is because this quantity is tightly constrained by Big Bang nucleosynthesis (BBN) light elements abundances observations (such as \cite{Schoneberg:2024ifp,PhysRevD.110.030001}) and therefore is rarely strongly perturbed in BAO analyses. Given the small range of variations typically considered, we show in \cref{fig:omegab} the results of varying $\Omega_b h^2$ between $0.02$ and $0.024$ (up to $\sim 4\sigma$ away from the central BBN prediction, even with the uncertainties adopted in \cite{Schoneberg:2024ifp} which conservatively encompass the different nuclear rate treatments), with the fiducial value of $\Omega_b h^2 \approx 0.022$ marked by a vertical line. With overall variations of only around $s \in [0.98, 1.02]$, it is clear that no large deviation among different ways of obtaining $s$ is expected. The relative shifts for $\alpha$  are shown in the right panel of \cref{fig:omegab}. Even for the complete DESI Y5 sample, the bias is only expected to become significant (reach $1/5$ of the statistical uncertainty) at deviations of $|\Delta (10^2 \Omega_b h^2)| \simeq 0.11$ (twice the conservative uncertainty of \cite{Schoneberg:2024ifp}). The required deviations to reach significance increase to $|\Delta (10^2 \Omega_b h^2)| \simeq 0.16$ for DESI~Y3 and $|\Delta (10^2 \Omega_b h^2)| \simeq 0.32$ for DESI~Y1.

Note that the simpler BAO-only method can give deviations of up to half a sigma; this is due to the accuracy of the de-wiggling method, not to the implementation of e.g., broadband correction terms. In particular, we have checked that with a more realistic broadband marginalization/minimization prescription, the performance is not improved.

\subsection{Varying \texorpdfstring{$\Omega_\mathrm{m}$}{Omegam}}\label{ssec:omegam}
BAO data are frequently used to constrain $\Omega_m$, as uncalibrated distance measurements primarily constrain the product $h r_d$ and the expansion rate $E(z) = H(z)/H_0$ \cite{desicollaboration2024desi2024vicosmological}, which in $\Lambda$CDM is determined by $\Omega_m$. It is therefore important to assess whether biases in $\Omega_m$ could arise solely from different definitions of the sound horizon. For fixed $H_0$\,, this is equivalent to testing the impact of $\Omega_m h^2$, and for fixed $\Omega_b h^2$, that of $\Omega_\mathrm{cdm}h^2$.

\begin{figure}
    \centering
    \includegraphics[width=0.48\linewidth]{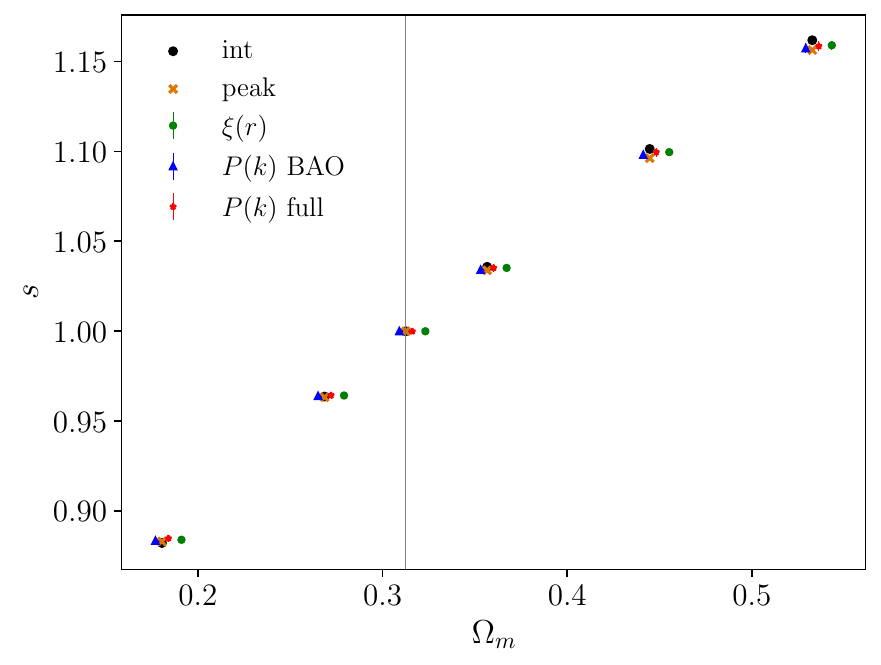}
    \includegraphics[width=0.48\linewidth]{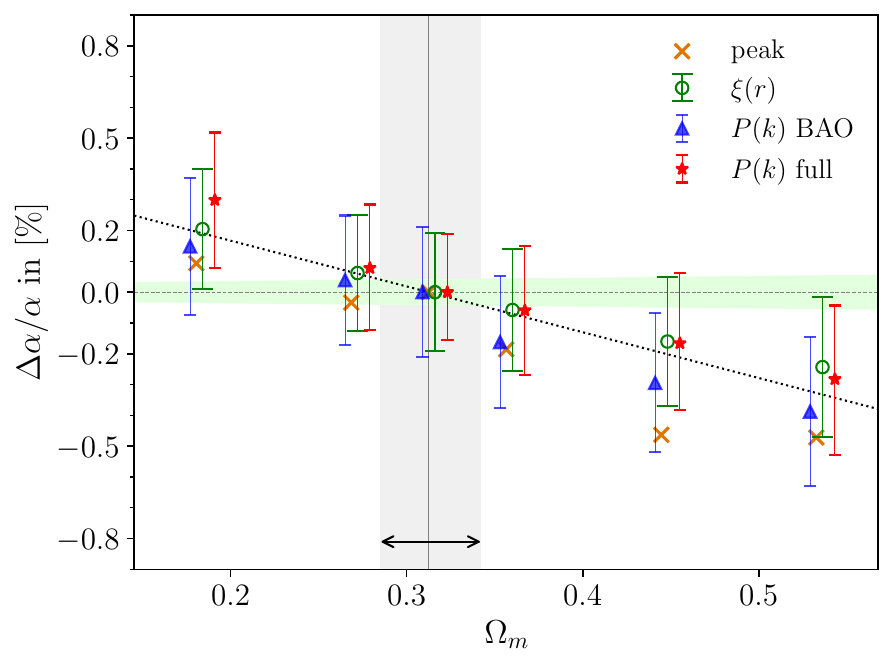}
    \caption{Same as \cref{fig:omegab} but for the matter fractional abundance $\Omega_m$.}
    \label{fig:omegam}
\end{figure}

In this case, a wide range of $\Omega_m$ values from 0.18 to 0.53 is considered (motivated by a range in $\Omega_\mathrm{cdm} h^2 \in [0.06, 0.22]$) inducing a much larger variation of $r_d^\mathrm{int}$ ($s \in [0.8,1.2]$, left panel of \cref{fig:omegam}). The relative shift in $\alpha$ values are shown in the right panel of \cref{fig:omegam}. With DESI~Y5 data, one can expect at most a $1\sigma$ bias between $r_d^\mathrm{int}$ and the other methods, even for such a large variation in $\Omega_m$. For the baseline $\Lambda$CDM analysis in \cite{DESI:2025zgx} of $\Omega_m = 0.2975 \pm 0.0086$, this does not represent a problem. However, when the parameter space describing the late Universe expansion is opened up, e.g. by introducing dynamical dark energy or curvature, $\Omega_m$ is not tightly constrained, and the resulting bias could potentially become significant. For DESI~Y5, this happens when the change from the fiducial reaches $|\Delta \Omega_m^\mathrm{Y5} | \approx 0.03$ (see \cref{tab:bias_analysis}). A careful investigation of this effect will be required when interpreting parameter constraints in extended cosmological models derived from the DESI-Y5 compressed parameters. For DESI~Y3 the threshold is $|\Delta \Omega_m^\mathrm{Y3}| \approx 0.043$, also potentially of interest in extended cosmologies.

For DESI~Y1 the bias is less important, only around half a sigma even at the extreme ends of the variation (see \cref{fig:varying_omega_m_desiy1lrg}):  the threshold of significance would be $|\Delta \Omega_m^\mathrm{Y1}| \approx 0.09$ (see also \cref{tab:bias_analysis}) well in the tail of the posterior distribution, even in the extended (thawing/curved) cosmologies investigated in \cite{desicollaboration2024desi2024vicosmological}. These numbers are obtained by linearly interpolating/extrapolating from the closest evaluated $x$ values. For all other considered surveys (smaller volumes and/or bigger noise), it is even less relevant. 

\subsection{Varying \texorpdfstring{$N_\mathrm{eff}$}{Neff}}\label{ssec:neff}
The effective number of neutrino species $N_\mathrm{eff}$ is defined as the ratio of radiation to photon energy densities. In the $\Lambda$CDM standard model, there are three neutrino species, contributing as $N_\mathrm{eff} = 3.044$ \cite{Drewes:2024wbw} (with small corrections from non-instantaneous decoupling and electroweak corrections). However, in cosmologies with additional dark radiation or dark sector equilibration (e.g., \cite{Berlin:2019pbq}), this quantity can be larger (or smaller) than in the standard model.

Variations of $N_{\rm eff}$ are known to have a strong impact on the sound horizon, while being weakly constrained by BAO data. Constraints are provided by BBN, except in scenarios where the dark radiation abundance is generated after the synthesis of the light elements. We therefore do not exclude very large or very small values \textit{a priori}, noting that current bounds from BBN and CMB are roughly at the level of $3 \pm 0.3$ for more standard cases \cite{Planck:2018vyg,Schoneberg:2024ifp}. However, note that recent results from Ref.~\cite{SPT-3G:2025bzu} hint at best-fit values of $\Delta N_\mathrm{eff} =-0.2 \pm 0.12$ with CMB data only, and $\Delta N_\mathrm{eff} = 0.64 \pm 0.42$ when including DESI+SPT data. 

\begin{figure}
    \centering
    \includegraphics[width=0.48\linewidth]{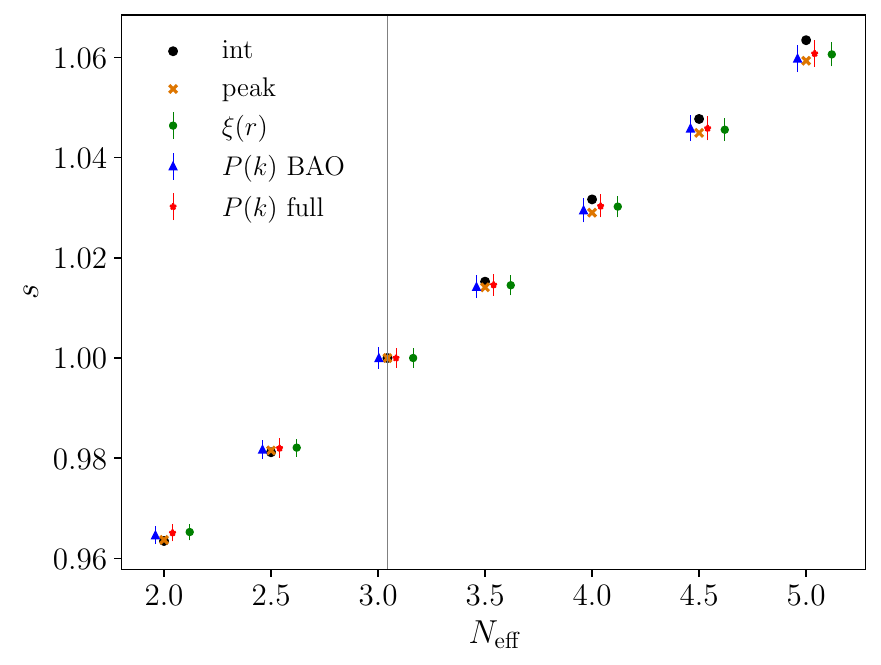}
    \includegraphics[width=0.48\linewidth]{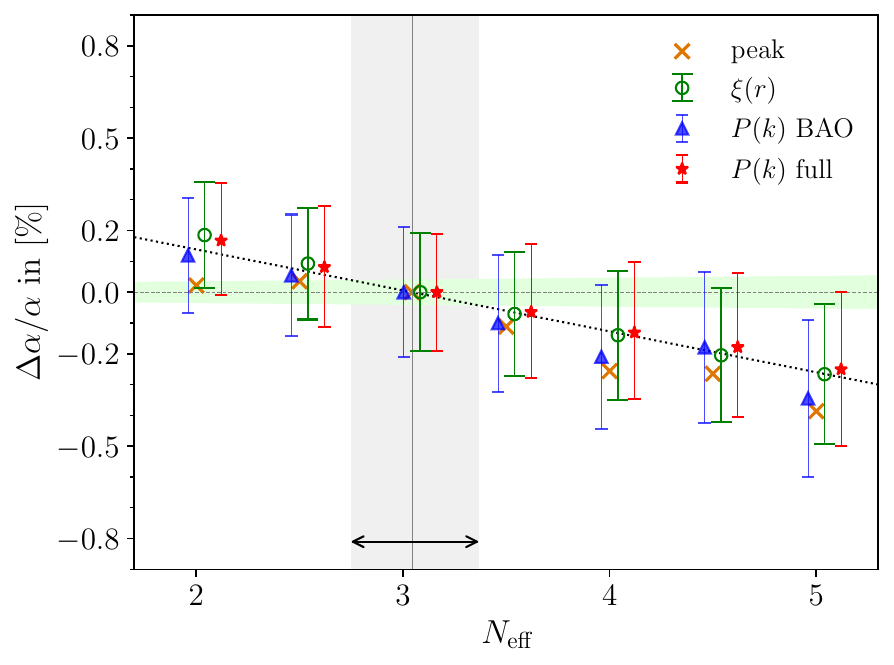}
    \caption{Same as \cref{fig:omegab} but for the effective number of neutrino species $N_\mathrm{eff}$. }
    \label{fig:neff}
\end{figure}

When varying $N_\mathrm{eff}$ in the range $N_\mathrm{eff} \in [2,5]$ --see \cref{ssec:compensated_cosmologies} for more discussion on how such extreme cases could arise in a BAO-only analysis-- we find values of $s \in [0.96,1.06]$ (\cref{fig:neff}, left panel). Over the full $N_{\rm eff}$ range, systematic shifts in $\alpha$ of up to $1\sigma$ can be expected for the DESI Y5 LRG survey specification, with all methods of estimating $s$ showing consistency. The threshold of significance for $x=N_{\rm eff}$ is $|\Delta x| \approx 0.3$. This suggests a possibly significant bias that would arise for DESI Y5 data when considering the CMB constraints from Ref.~\cite{SPT-3G:2025bzu} discussed above.

Even if these deviations are constrained by BBN and CMB data in scenarios with \textit{ordinary} free-streaming dark radiation, there are extended dark radiation models in which such bounds don't apply: {\it i)} When the dark radiation abundance is created only after BBN, most of the BBN bounds relax, and larger abundances are in principle possible. {\it ii)} In particular, when the dark radiation is not free-streaming, the CMB limits weaken significantly. This is the case for self-interacting dark radiation, Wess-Zumino dark radiation, Majoron dark radiation, and other similar models, see for example \cite[Sec.~3.2]{Schoneberg:2021qvd} or \cite[Sec.~7]{DiValentino:2021izs}.

\subsubsection{A note on Helium}\label{ssec:Helium}

The sound horizon is modified by the primordial Helium abundance, $Y_\mathrm{He}$, through \cref{eq:ne}: the Helium abundance changes the ionization fraction, which in turn changes the time of baryon drag. Therefore, whether the Helium fraction is assumed to be fixed (as throughout this manuscript) or allowed to vary according to BBN constraints can make a small difference, in principle. In practice the dependence as been shown in  \cite{Schoneberg:2021uak} to be weak: $r^\mathrm{int}_d \propto Y_\mathrm{He}^{0.008}$. 

We find numerically that the differences between accounting for and not accounting for the BBN Helium abundance variations are negligible. For example, the slope of the bias changes from $-0.00133 \pm 0.00004$ when not accounting for changes in $Y_\mathrm{He}$ (c.f. \cref{tab:bias_analysis}) to $-0.00138 \pm 0.00004$ when using \texttt{CLASS} to convert a given $\{N_\mathrm{eff},\Omega_b h^2\}$ into a value of $Y_\mathrm{He}$\,. This small difference is within the numerical uncertainty of our analysis and therefore we do not consider this effect significant. Similar statements hold for the corresponding maximum parameter deviations $\Delta x_\pm$, which are even tighter when the $Y_\mathrm{He}$ effect is taken into account, thereby rendering our baseline analysis conservative.

\subsection{Shift of recombination}\label{ssec:me}

A shift of the recombination redshift has a direct impact on the sound horizon. To investigate the effect, we employ a toy model for which the electron mass in the early universe $m_e^\mathrm{early}$ before recombination differs from the one measured locally in laboratories ($m_e^\mathrm{late}\simeq 511\mathrm{keV}$) by $\pm 20\%$, a wider range than currently accepted variations for the electron mass (see \cite[Table~1]{Schoneberg:2024ynd}). This model has been shown to successfully alleviate the Hubble tension in previous studies (see \cite{Schoneberg:2021qvd,Schoneberg:2024ynd}).

\begin{figure}
    \centering
    \includegraphics[width=0.48\linewidth]{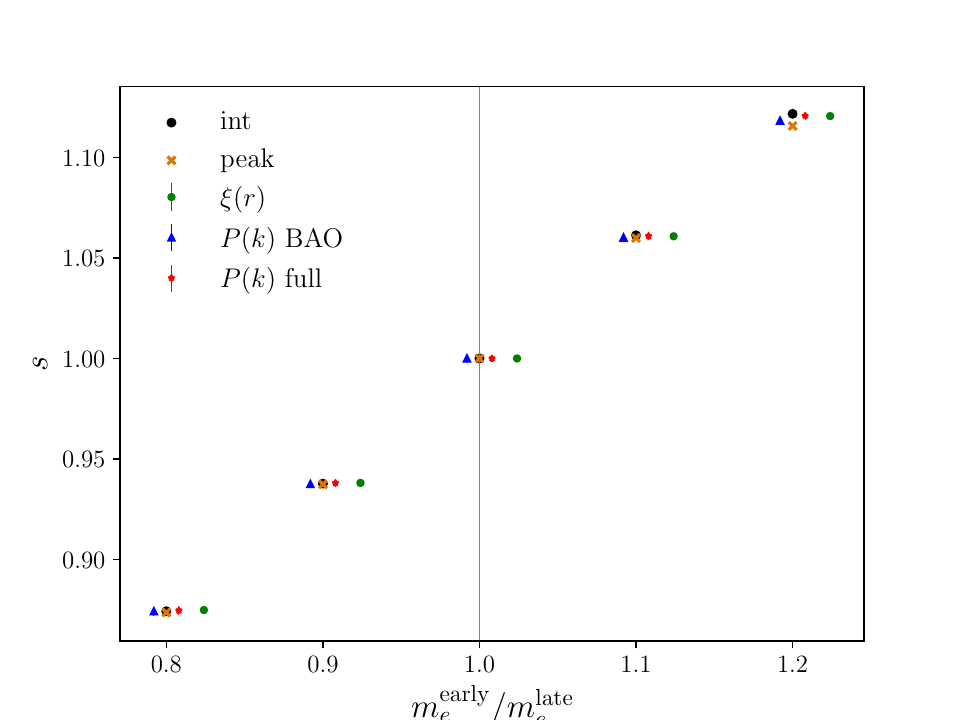}
    \includegraphics[width=0.48\linewidth]{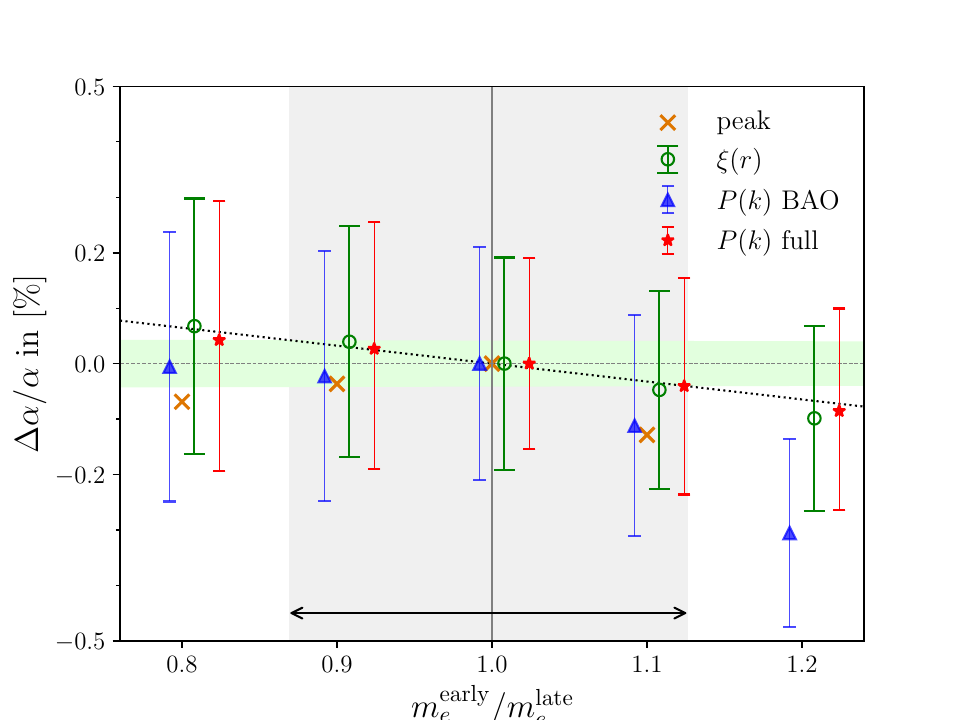}
    \caption{ Same as \cref{fig:omegab} but for various values of the electron mass in the early universe $m_e^\mathrm{early}$ different from its laboratory value $m_e^\mathrm{late}$\,.}
    \label{fig:varying_me}
\end{figure}

As summarized in \cref{fig:varying_me}, we find no significant deviations even for extremely large variations in the electron mass in the early Universe, demonstrating that the BAO standard ruler is robust to changes in the sound horizon arising from this class of early-Universe models.
The corresponding $m_e^{\rm early}$ deviations required for significant bias shown in \cref{tab:bias_analysis} are far beyond current limits from the CMB \cite{SPT-3G:2025bzu} and will remain so even for DESI-Y5 data.

\subsection{Shifted sound horizon through pre-recombination dark energy}\label{ssec:ede}

\begin{figure}
    \centering
    \includegraphics[width=0.48\linewidth]{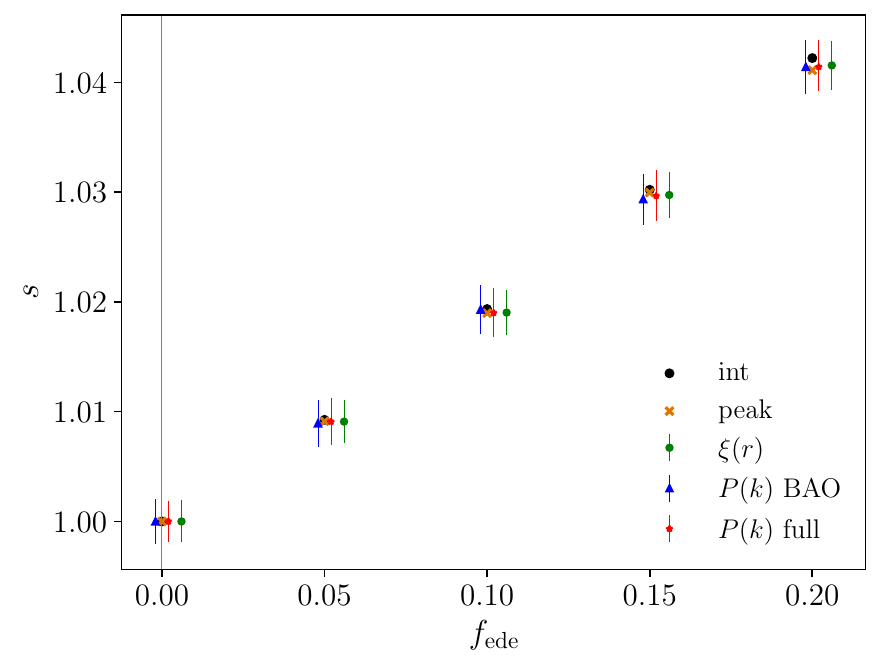}
    \includegraphics[width=0.48\linewidth]{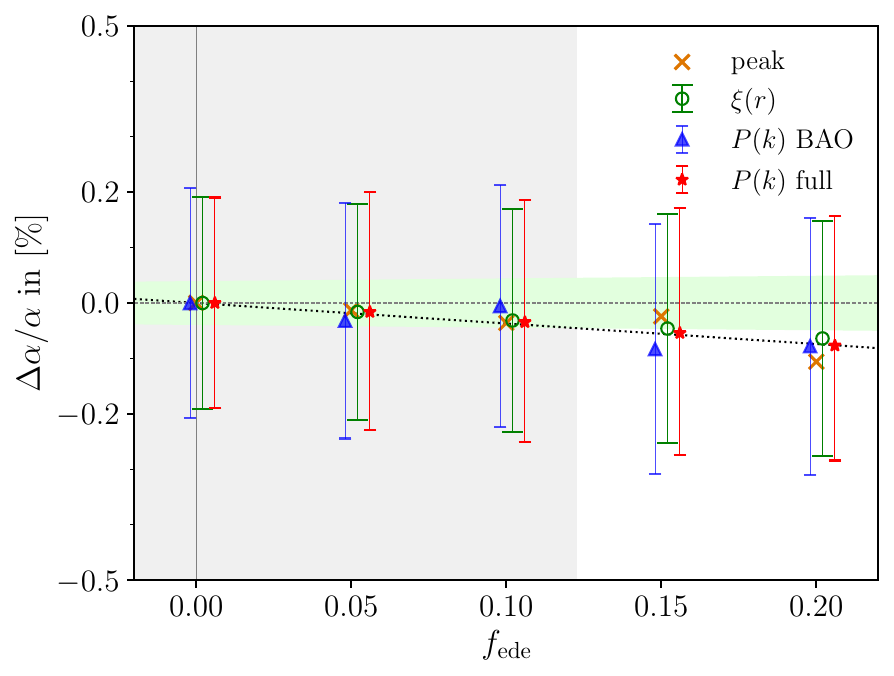}
    \caption{Same as figure \ref{fig:neff} but for  the  $f_\mathrm{ede}$ parameter of early dark energy.}
    \label{fig:varying_fede}
\end{figure}

Similarly to the shift of the recombination redshift, another way to strongly influence the sound horizon is by including additional contributions to the Hubble rate in the early universe. This can be accomplished, for example, using early dark energy, see \cite{Poulin:2018cxd,Schoneberg:2021qvd} for the impact on the sound horizon and Hubble tension and \cite{Poulin:2023lkg} for a review of this model. The fraction of early dark energy, the model parameter $f_\mathrm{ede}$, quantifies how strongly the model shifts the sound horizon.\footnote{The function  $\Omega_\mathrm{ede}(z)$ is the early dark energy density in units of the critical density as a function of redshift throughout the cosmic history, and $f_\mathrm{ede}$ is the maximum value of this function.}  As shown in \cref{fig:varying_fede}, the bias is negligible for all methods. Even for DESI-Y5 data, the parameter shift required to give a significant bias is excluded by current CMB bounds ($f_{\rm ede}\lesssim 0.1$ at 95\% confidence) \cite{SPT-3G:2025bzu}.

\subsection{Note on other parameter variations}\label{ssec:other_params}

Other commonly considered variations of the $\Lambda$CDM model do not have a large impact on the sound horizon. For example, in the case of dynamical dark energy, the power spectrum is overall suppressed due to the presence of dark energy, and the individual cosmological distances $D_H,\,D_M,\,D_V$ are impacted, but the sound horizon remains the same (see also the discussion in \cref{ssec:bao_int}). Cases that do not affect the sound horizon are, in principle, irrelevant for this study. These include, for example, dynamical dark energy and curvature, as well as many other late-time modifications. However, as we argued above, such variations can be important if such parameters open up degeneracies with other parameters (e.g., $\Omega_m$), allowing larger variations than previously considered.

Two cases are of particular interest: the Hubble constant $H_0$ and the sum of neutrino masses, $m_\nu$\,. While these parameters do not directly affect the sound horizon (in our parameterization), they are degenerate with other parameters that do.

$H_0$ does not impact $r_d$ directly because the integrand in \cref{eq:rd_int} in flat $\Lambda$CDM depends only on $\Omega_x h^2$ with $x \in \{\gamma, r, b, m\}$ (photons, radiation, baryons, total matter) via the Hubble parameter. The photon/radiation physical densities ($\Omega_\gamma h^2$ and $\Omega_r h^2$) are measured from the CMB, and the baryon physical density ($\Omega_b h^2$) from the light element abundances in BBN as well as from the CMB. However, it is the fractional matter density parameter $\Omega_m$ that is tightly measured e.g., from BAO themselves, from supernovae of type Ia, \cite{Brout:2022vxf}, or cosmic chronometers \cite{Moresco:2023zys}; in this case $H_0 = h \cdot 100\mathrm{km/s/Mpc}$ has an indirect impact on the sound horizon -- for a given fixed $\Omega_m$ it changes $\Omega_m h^2$. Since the sound horizon depends primarily on $\Omega_m h^2$, the results of \cref{ssec:omegam} can be recast in terms of $h$ as well.\footnote{In practice one would compute the shift in $\Delta \Omega_m$ that would cause the same change as a given $\Delta h$ by equating $\Delta \Omega_m (h^\mathrm{fid})^2 = \Delta (\Omega_m h^2) \approx 2 \Omega_m^\mathrm{fid} h^\mathrm{fid} \Delta h$. This allows mapping a change $\Delta h$ into a change $\Delta \Omega_m$ that can be compared to the results of \cref{ssec:omegam}. The range of variations in $\Delta h$ required for significant bias -- e.g. $|\Delta h| \approx 0.03$ for DESI-Y5 (total) -- is at the same level as the Hubble tension.}

\begin{figure}
    \centering
    \includegraphics[width=0.48\linewidth]{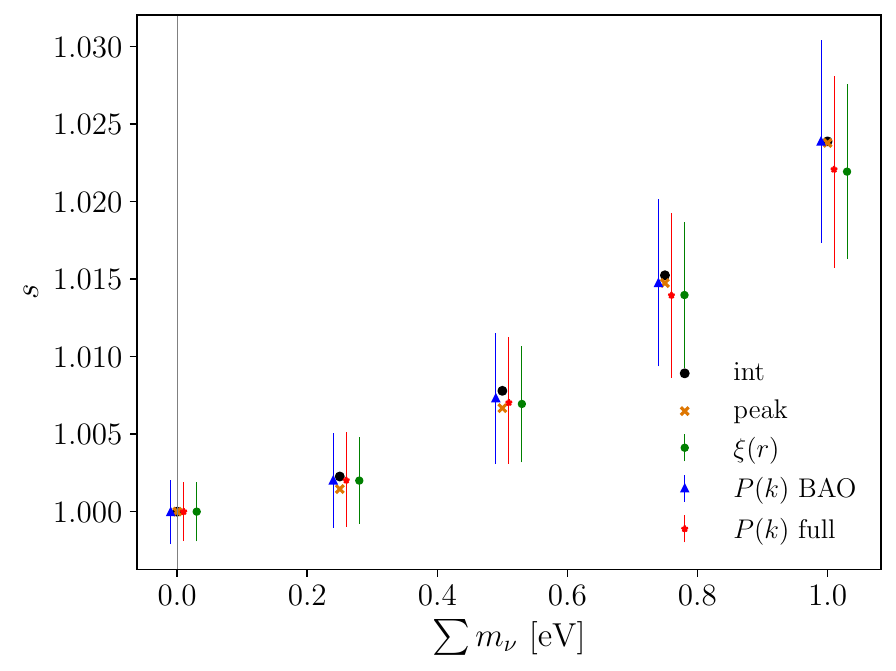}
    \includegraphics[width=0.48\linewidth]{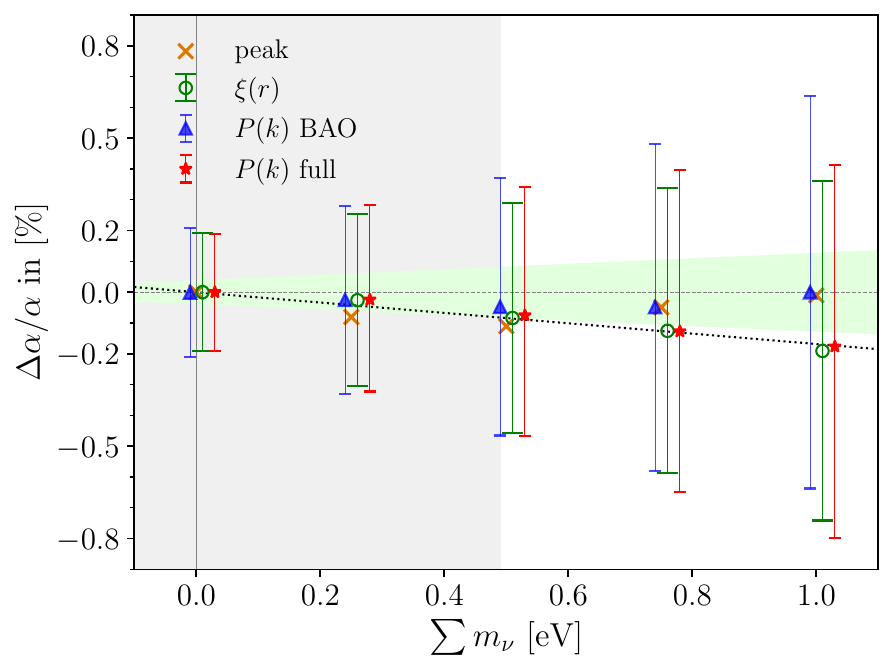}
    \caption{As for \cref{fig:omegam} but for 
    the total sum neutrino mass $\sum m_\nu$ in eV. }
    \label{fig:mnu}
\end{figure}

For sufficiently small values, the neutrino mass ($m_\nu$, or equivalently $\sum m_\nu$) does not affect the sound horizon (for a fixed $\Omega_\mathrm{cdm}h^2+\Omega_b h^2$). However, for large enough masses, neutrinos become non-relativistic already at the drag epoch. In those cases, the additional contribution to $H(z)$ in the denominator of \cref{eq:rd_int} becomes relevant, leading to a per cent-level decrease of the sound horizon (and, therefore, a per cent-level increase in $s$). We investigate this effect in a range of $\sum m_\nu$ that includes the range allowed by KATRIN \cite{doi:10.1126/science.adq9592} ($m_\nu<0.45$ eV, 90\% CL) and show it in \cref{fig:mnu}. We note that, for DESI Y5 data, no significant biases are observed for $\sum m_\nu<$0.5 eV (see also \cref{tab:bias_analysis}).

\subsection{Special (compensated) cosmologies}\label{ssec:special}
In \cite{Sanz-Wuhl:2024uvi} two special cosmologies (Cosmology$+$ and Cosmology$-$, see \cref{tab:cosmo_plusminus}) were studied whereby relatively large deviations of the curvature, $\Omega_m$ and $H_0$ parameters values from the fiducial combine to keep $\alpha_\parallel(z)$, $\alpha_\perp(z)$ approximately constant within a certain redshift range (hence the name {\it compensated}). As a result, despite their seemingly extreme values for the cosmological parameters,  Cosmology$+$ and Cosmology$-$ are not excluded by galaxy and quasar BAO data alone (excluding Lyman-$\alpha$ BAO).

\begin{figure}
    \centering
    \includegraphics[width=0.48\linewidth]{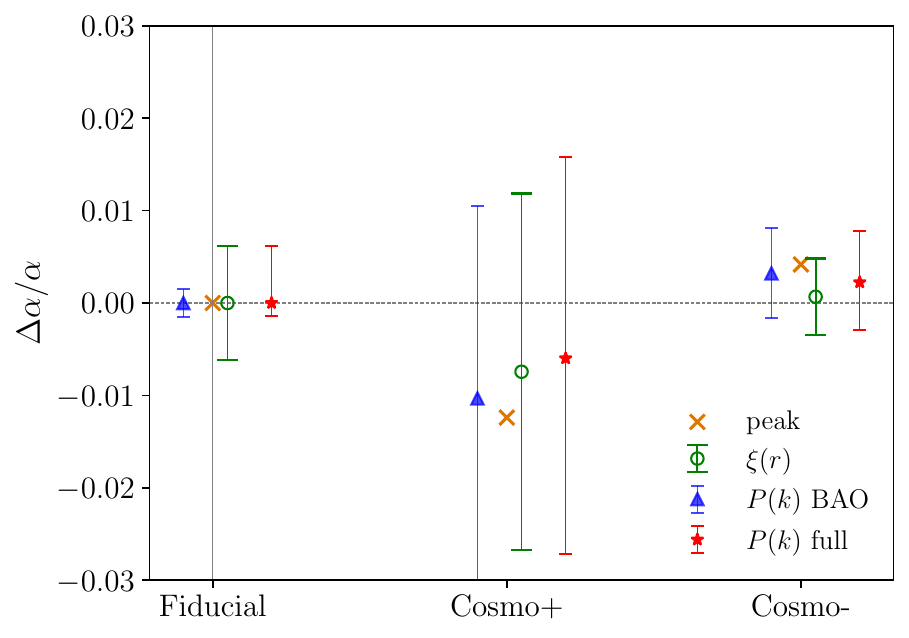}
    \includegraphics[width=0.48\linewidth]{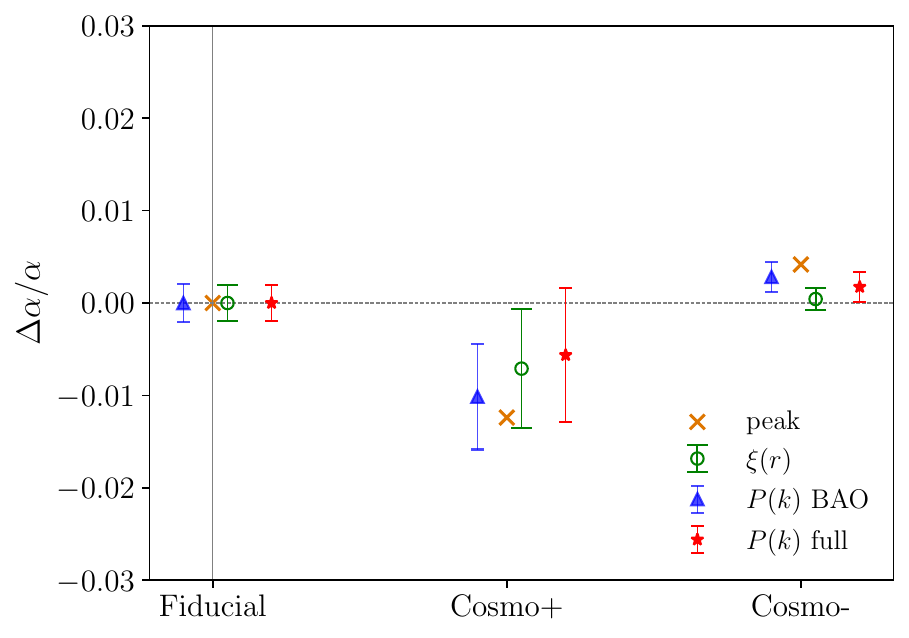}
    \caption{As for the right panel of \cref{fig:omegab}, but using the cosmologies of \cref{tab:cosmo_plusminus}. We show the results for DESI Y1 (left) and DESI Y5 (right).}
    \label{fig:special}
\end{figure}

In \cref{fig:special} we show the bias $\Delta \alpha/\alpha$ for these two cosmologies. The bias is around $-0.7\%$ for Cosmology+, and $+0.1\%$ for Cosmology- \footnote{Ref~\cite{Sanz-Wuhl:2024uvi} adopted their fiducial model to the shifted cosmology and then computed the difference to the original Fiducial* cosmology, leading to an inversion of the sign. Further, they used our quick estimator ($\alpha^\mathrm{peak}$) and therefore cite 1.3\% and $-0.1\%$, respectively.}, corresponding to $\sim 0.5\sigma$ for DESI Y1 and $\sim 1\sigma$ level for DESI Y5. Note that the different sizes of the uncertainties on the $\Delta \alpha/\alpha$ visible in \cref{fig:special} are due to two effects; i) the power spectrum amplitude for the Cosmology+ is much smaller than than for Cosmology- due to the small value of $\Omega_m h^2$, leading to large uncertainties (c.f. \cref{fig:A_s} for the scaling of uncertainty with power spectrum amplitude) and ii) the uncertainties shown are relative, and since Cosmology+ happens to have a small value of $s$ the relative error bars are enhanced. Simultaneously adjusting $A_s$ would lead to results where the uncertainties are closer in size.

\begin{table}
    \centering
    \begin{tabular}{|c|c c c c|}
    \hline
        Cosmology & $\Omega_k$ & $\Omega_m$ & $\Omega_b h^2$ & $h$  \\ \hline \hline
        Fiducial* & $0$ & 0.31 & 0.02189 & 0.676 \\
        Cosmology+ & $0.15$ & $0.242$ &  $0.0095$ &  $0.505$ \\
        Cosmology- & $-0.15$ & $0.336$ & $0.052133$ & $1$\\
        \hline
    \end{tabular}
    \caption{Parameters of the Cosmology+ and Cosmology$-$ from \cite{Sanz-Wuhl:2024uvi}. The models are curved $\Lambda$CDM cosmologies. The model includes a single neutrino with a mass of 0.06eV. The starred fiducial cosmology is that of \cite{Sanz-Wuhl:2024uvi}.}
    \label{tab:cosmo_plusminus}
\end{table}

\subsection{Fully compensated cosmologies}\label{ssec:compensated_cosmologies}

The cosmologies of \cref{ssec:special} may be regarded as extreme cases, already ruled out by other observations such as CMB, BBN, or local $H_0$ measurements \cite{Riess:2021jrx}. We show below other compensated cases, much harder to rule out,  where the potential bias in $\alpha$ is non-negligible. In this case a change in $N_\mathrm{eff}$ can be perfectly compensated by a change in $h$ for fixed $\Omega_m$ and $\Omega_b h^2$, as to keep $\alpha$ unchanged over a broad redshift interval; $H_0$ remains broadly consistent with current constraints and other probes, such as SNe Ia or cosmic chronometers measuring $\Omega_m$\,, remain unaffected. 

We again adopt the fiducial cosmology of \cref{eq:fiducial}, and the cosmological parameters that provide such an effect for two cases are detailed in \cref{tab:cosmo_compensated}.

\begin{table}
    \centering
    \begin{tabular}{|c|c c c| c c|}
    \hline
        Cosmology & $N_{\rm eff}$ & $\omega_{cdm}$ & $h$   & $\Omega_m$ & $\Omega_b h^2$  \\ \hline \hline
        Fiducial  & $3.046$       & $0.1196 $      & $0.674$ & 0.312 & 0.02207\\
        Comp. 1 (high $N_\mathrm{eff}$)   & $5$           & $0.1583$       &  $0.760$ & 0.312 & 0.02207\\
        Comp. 2 (low $N_\mathrm{eff}$)  & $1$           & $0.0796$       & $0.571$ & 0.312 & 0.02207 \\
        \hline
    \end{tabular}
    \caption{Parameters of the two compensated cosmologies, one with a larger $N_{\rm eff}$ value and one with a smaller $N_{\rm eff}$ value compared to the fiducial cosmology, as well as a fiducial Planck2018-based cosmology. The values have been chosen such that $\Omega_m$ and $\Omega_b h^2$ remain consistent with the Planck-determined values (and therefore with BBN, uncalibrated BAO and SNe Ia data).
    The Helium abundance $Y_\mathrm{He}$ has been fixed to 0.24, see also \cref{ssec:Helium}.}
    \label{tab:cosmo_compensated}
\end{table}

\begin{figure}
    \centering
    \includegraphics[width=0.48\linewidth]{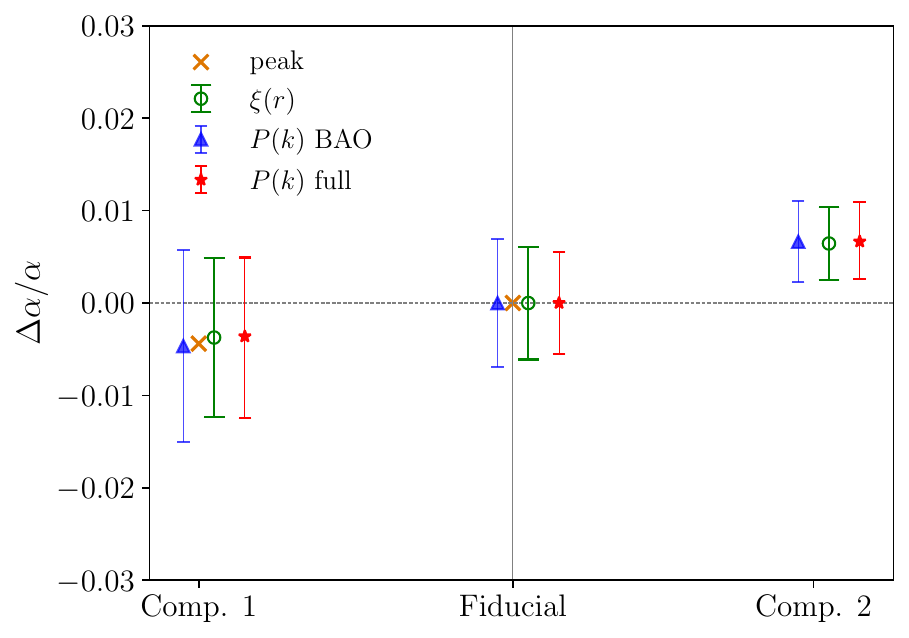}
    \includegraphics[width=0.48\linewidth]{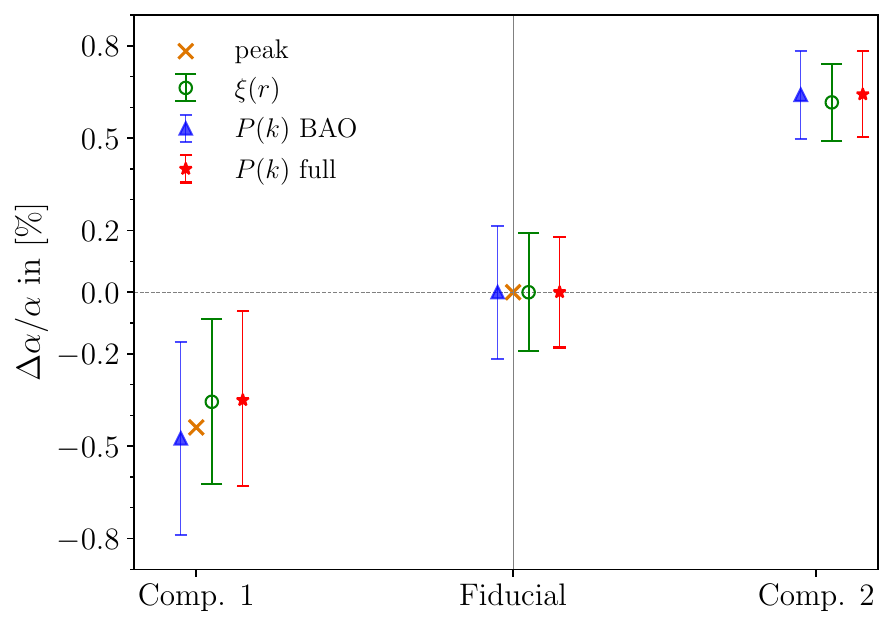}
    \caption{Same as \cref{fig:special}, but for the cosmologies in \cref{tab:cosmo_compensated} instead. Note the different scales of the two axes. (Left) DESI Y1 (total) survey, (Right) DESI Y5 (total) survey.}
    \label{fig:degeneracy}
\end{figure}

We show the resulting $\Delta\alpha/\alpha$ in \cref{fig:degeneracy}. Significant biases can be observed which reach beyond the $1\sigma$ level for DESI Y5 survey specifications, despite $\alpha^{\rm int}(z)=1$ at all redshifts by construction.
While both CMB and BBN rule out the extreme values of $\Delta N_\mathrm{eff}$ explored in the figure, as seen in \cref{ssec:neff}, in this fully compensated direction, the parameter deviations required for a bias of 1/5 of the size of $\alpha$'s statistical uncertainty is only
$|\Delta N_\mathrm{eff}^{\rm  Y5}| \approx 0.17$, $\Delta |N_\mathrm{eff}^\mathrm{Y3}| \approx 0.24$, and $\Delta |N_\mathrm{eff}^\mathrm{Y1}| \approx 0.5$. These results are more stringent than the ones in \cref{ssec:neff} due to the impact of $h$ (see \cref{ssec:other_params}), thus showing that such compensated directions must be accounted for when considering whether a given parameter difference is considered \enquote{reasonable} or not.

\subsection{Summary}

\begin{table}[ht!]
    \centering

    \resizebox{\columnwidth}{!}{

    \begin{tabular}{|c|c c | c c| c|}
    \hline
        & \multicolumn{2}{|c|}{DESI-Y5 (total)} & \multicolumn{2}{|c|}{DESI-Y1 (total)} & \\
       Parameter  & $\Delta x_-$& $\Delta x_+$  & $\Delta x_-$ & $\Delta x_+$ & Slope of the bias \\ \hline \hline
       \rule{0pt}{2.4ex}
$10^2 \Omega_b h^2$ & $-0.1315 ^{+ 0.0042 }_{- 0.0047}$ & $0.1071 ^{+ 0.0031 }_{- 0.0028}$ & $-0.463 ^{+ 0.016 }_{- 0.017}$ & $0.3197 ^{+ 0.0081 }_{- 0.0077}$ & $34.5 \pm 1.05$\\
$\Omega_m$ & $-0.0288 ^{+ 0.0023 }_{- 0.0027}$ & $0.0294 ^{+ 0.0029 }_{- 0.0024}$ & $-0.0901 ^{+ 0.0068 }_{- 0.0084}$ & $0.0956 ^{+ 0.0095 }_{- 0.0077}$ & $-0.0149 \pm 0.0013$\\
$N_\mathrm{eff}$ & $-0.2988 ^{+ 0.0086 }_{- 0.0094}$ & $0.321 ^{+ 0.011 }_{- 0.010}$ & $-0.831 ^{+ 0.019 }_{- 0.019}$ & $1.011 ^{+ 0.029 }_{- 0.028}$ & $-0.00133 \pm 0.00004$\\
$m^\mathrm{early}_e/m^\mathrm{late}_e$ & $-0.140 ^{+ 0.012 }_{- 0.014}$ & $0.1198 ^{+ 0.0098 }_{- 0.0087}$ & $-0.484 ^{+ 0.043 }_{- 0.051}$ & $0.290 ^{+ 0.018 }_{- 0.016}$ & $-0.00323 \pm 0.00027$\\
$f_\mathrm{ede}$ & $-0.1070 ^{+ 0.0024 }_{- 0.0026}$ & $0.1208 ^{+ 0.0033 }_{- 0.0030}$ & $-0.2857 ^{+ 0.0062 }_{- 0.0063}$ & $0.361 ^{+ 0.010 }_{- 0.010}$ & $-0.0037 \pm 0.0001$\\
$\sum m_\nu$ [eV] & $-0.1829 ^{+ 0.0061 }_{- 0.0064}$ & $0.482 ^{+ 0.047 }_{- 0.040}$ & $-0.2973 ^{+ 0.0054 }_{- 0.0058}$ & $-1.500 ^{+ 0.132 }_{- 0.153}$ & $-0.00168 \pm 0.00008$\\
     \hline
     \end{tabular}
     }
    \caption{Maximum allowed deviations for the cosmological parameters such that the corresponding bias remains below the critical threshold of $1/5$-th of the statistical uncertainty on $\alpha$, as well as the slope of the corresponding linear bias-parameter relation. The parameters $\Delta x_{\pm}$ are defined as in the text. The uncertainties are numerical and are propagated from the linear bias slope fit.
    If there is no value cited, it either means that parameter variations in this direction are fundamentally disallowed or that they could not be determined with the present methodology due to small slopes of the linear bias relation.  These values have been obtained with the $P(k)~\mathrm{full}$ method, though the $\xi(r)$ method returns results that are compatible within the cited numerical uncertainties.}
    \label{tab:bias_analysis}
\end{table}

For single parameters and for the $P(k)$ full method, the slope of the bias relation together with the parameter deviation required to reach a bias as large as 1/5 of the statistical uncertainty are reported in \cref{tab:bias_analysis}. The parameter deviation is denoted as $\Delta x_\pm$ in the positive/negative direction of the parameter, respectively.\footnote{We linearly extrapolate the size of the statistical uncertainties beyond our chosen parameter ranges when $\Delta x_\pm$ is larger than this parameter range.} The uncertainties are numerical and are propagated from this linear bias slope fit.\footnote{The bias as a function of the parameter value does not typically lie in a perfect line, both due to quadratic corrections and due to numerical noise in the minimization procedure determining the mean value of the bias for a given parameter value. We have found the former to be negligible in these cited cases, while the latter can be used to derive numerical uncertainties.}

For individual parameter variations within usually adopted external priors, the effect is generally small except for $\Omega_m$ and especially $N_{\rm eff}$, where care must be taken when degeneracies with other beyond $\Lambda$CDM parameters extend the viable range. In a high-dimensional parameter exploration, when combinations of parameters are allowed to deviate from the fiducial values, the effect can be important. In particular, we have shown that there are compensated models involving variations in $N_\mathrm{eff}$ where biases may be relevant for DESI-Y5 data. 

We show in \cref{sec:nonlin_compare} that this conclusion does not change when considering non-linearities, redshift space distortions and a more realistic analysis pipeline. While the detailed and quantitative assessment of the effects of reconstruction is beyond the scope of this paper, the analysis of \cite{Carter_2020} indicates that reconstruction is not expected to alter the conclusions in any significant way,
see also the discussion in \cref{sec:forecast}.

\section{Nonlinearities and realistic analysis pipeline}\label{sec:nonlin_compare}

Given our somewhat simplistic analysis of the previous sections, here we check whether the effects seen in \cref{sec:results} persist both qualitatively and quantitatively in a more realistic analysis. We adopt the official DESI pipeline products, closely matching the power spectrum model of \cite{DESI:2024uvr}, but lacking mainly the effects of reconstruction, masking, and the more complicated true covariance matrix. This provides a setup as close as possible to the real data analysis, without introducing unnecessary complications in forecasting the Y5 case and without having to resort to expensive (and numerically statistical) N-body simulations to model reconstruction effects. For the modeling of how the sound horizon mismatch might be impacted by these effects, see \cite{Carter_2020} and the discussion in \cref{sec:forecast}.

As it will be clear later, it will suffice to limit ourselves to the cosmologies of \cref{ssec:compensated_cosmologies} for this comparison. We follow the methodology described in \cref{ssec:desi_bao_analysis_method} for a redshift $z=0.704$ (effective redshift $z_{\rm eff}$ of DESI~Y1 LRG2 sample \cite{desi_redshifts_tr6y-kpc6}) with the volume and shot noise of the DESI Y5 LRG sample (see \cref{tab:surveys}). The priors used for the MCMC inference follow common practice (e.g., \cite{DESI2024III_Adame_2025}) and are detailed in \cref{tab:priors_compensated_cosmologies}.

\begin{table}
    \centering
    \begin{tabular}{|c|ccccccc|}
    \hline
         Parameter  &  $q_{\rm iso}$&$q_{\rm AP}$  & $b_1$ & $d\beta$ & $\Sigma_s$ & $\Sigma_\parallel$ & $\Sigma_\perp$\\ \hline \hline
         Prior      & $\mathcal{U}[0.8, 1.4]$ & $\mathcal{U}[0.8, 1.2]$  & $\mathcal{U}[0, 3]$  & $\mathcal{U}[0.7, 1.3]$  & $\mathcal{N}(*, 2)$ & $\mathcal{N}(*, 2)$ & $\mathcal{N}(*, 2)$\\
         \hline
    \end{tabular}
    \caption{Priors used for the MCMC of \cref{sec:nonlin_compare}. Wherever we show a star ($*$), it means that we have used the one obtained from the minimization as the central value.}
    \label{tab:priors_compensated_cosmologies}
\end{table}

To validate the pipeline, we test it using a mock data vector generated from the \textit{linear} power spectrum of the models of \cref{ssec:compensated_cosmologies}, \cref{tab:cosmo_compensated}, including redshift space distortions. We obtain $\Delta\alpha/\alpha = 0.007\pm0.002$ for the cosmology with lower $N_{\rm eff}$ (Comp. 2) and $\Delta\alpha/\alpha = -0.004\pm0.003$ for the cosmology with higher $N_{\rm eff}$\,(Comp. 1), in very good agreement with the values shown in \cref{fig:degeneracy}.

When turning to mock data generated using a fully non-linear prescription (see \cref{ssec:desi_bao_analysis_method}), we expect an additional bias, $\delta \alpha^\mathrm{NL}$, caused by non-linearities displacing the BAO signal to slightly higher $k$, which is not modeled directly by the BAO template used in the fit. This additional bias, which is present even if $r_d^{\rm int}\equiv r_d^{\rm obs}$, in a real data analysis would be mitigated by reconstruction algorithms and is therefore not a focus of this study. We proceed to quantify it (and then correct for it) as shown in \cref{appendix:quantify_nl_bao_shift}.

\begin{table}
    \centering
    \begin{tabular}{|c|c c|}
    \hline
        Cosmology                                               & Comp. 2 (low $N_{\rm eff}$) & Comp. 1 (high $N_{\rm eff}$) \\ \hline \hline \rule{0pt}{2.5ex}
        $s^{P(k)\,{\rm full}}$                              & $0.852\pm 0.002$  & $1.125\pm 0.003$   \\
        $s^{\rm DESI, lin}$                                     & $0.853\pm0.002$ & $1.125\pm0.003$   \\ 
        $s^{\rm DESI, non-lin}$                                 & $0.858\pm 0.0044$  & $1.128\pm 0.006$   \\  \hline \rule{0pt}{2.5ex}
        $\Delta \alpha^{P(k)\,\mathrm{full}}/\alpha$ (in $\%$)             & $0.64\pm 0.16$      & $-0.35\pm 0.31$       \\
        $\Delta \alpha^{\rm DESI, lin}/\alpha$ (in $\%$)             & $0.7\pm 0.2$      & $-0.4\pm 0.3$       \\
        $\Delta \alpha^{\rm DESI, non-lin}/\alpha$ (in $\%$)  & $0.8\pm 0.5$      & $-0.4\pm 0.5$       \\\hline
    \end{tabular}
    \caption{Values of $s$ and systematic biases in $\alpha$ for the two compensated cosmologies, one with a larger $N_{\rm eff}$ value and one with a smaller $N_{\rm eff}$ value than the fiducial cosmology, as defined in \cref{tab:cosmo_compensated}. The results from \cref{ssec:compensated_cosmologies} and \cref{sec:nonlin_compare} (titled here \enquote{DESI, lin} and \enquote{DESI, non-lin} agree nicely.}
    \label{tab:alpha_compensated_desi_pipeline}
\end{table}

The resulting systematic shifts $\Delta \alpha ^{\rm DESI, non-lin.}/\alpha$ are $0.8\pm 0.5 \%$ for the lower $N_{\rm eff}$ case (Comp.2) and $-0.4\pm 0.5$ for the higher $N_{\rm eff}$ case (Comp.1), as shown in \cref{tab:alpha_compensated_desi_pipeline}. This is in good agreement with the $\Delta\alpha/\alpha$ values reported in \cref{sec:results}, but with larger error-bars. While our way to remove the non-linear BAO shift may be approximated, it is important to recall that 
since re-construction significantly reduces the uncertainty in the estimation of $\alpha$ (without reducing the size of this sound horizon mismatch effect) by around a factor of $\sim 1.5-2.5$ (compare e.g. \cite[Tab.~16]{DESI:2024uvr}), we expect the smaller uncertainties obtained in the linear case to be the ones that are more representative of the final uncertainty in a full DESI analysis including reconstruction.

In essence, the mismatch of the sound horizon (and thus the induced systematic in $\alpha$  computed under the linear approximation) holds when including non-linearities. Therefore, especially in high-dimensional parameter exploration, when $N_{\rm eff}$ and $\Omega_m$ can deviate significantly from their fiducial values, this systematic may become important and cannot be ignored.

\section{Correcting for this effect and the bias}\label{sec:correcting}
Here, we outline several possible methods for correcting this effect. We envisage that the suitability of each solution will depend on the specific context or application.

\paragraph{Avoid making the $s^{\rm int}=s^{\rm obs}$ approximation.} This is, for example, the case of the full modeling approach that does not involve computing the compressed parameters $\alpha$. It has, however, the disadvantages of being computationally very slow and expensive and of losing the model independence of the compressed variables approach. While the first disadvantage could be mitigated by resorting to fast emulators, the model dependence remains, and different emulators have to be trained for each different model. An intermediate solution is to compute the bias as in this work using a high-precision limit\footnote{In principle, the correction should be computed using the specific parameters of each survey. In fact, the relative weights of the different scales are survey-dependent. Nevertheless, the results for different survey specifications (such as those from \cref{tab:surveys}) are very consistent to within the numerical accuracy of the minimization. This consistency suggests that the correction could be evaluated for a high-precision idealized survey (e.g. \enquote{HUGE} from \cref{tab:surveys}) and applied to realistic surveys, largely independent of survey specifics, subject to further validation.} (see e.g. \cref{sec:forecast}) for each cosmological parameter point of an MCMC exploration; while this is a somewhat expensive approach, due to the many simplifications of the likelihood, it should still be faster than the corresponding full modeling analysis (requiring no EFT or other non-linear corrections, requiring no power spectrum multipoles, being applicable for different tracers/redshifts). Additionally, for a fixed value of $s$ \cref{eq:likelihood_full} actually reduces to a weighted linear least squares fit for the parameters $a_i$ and $B$, which can be solved analytically.\footnote{Grouping $\theta = (B, a_1, \ldots a_5)$ and $f(k,s) = (P^\mathrm{fid}(ks), (ks)^2, \ldots, (ks)^{-2})$ and $W(k)=V k^2/(4 \pi^2 (\mathcal{N}+ P^\mathrm{fid}(k))^2)$ we can construct the vector $D = \int W(k) P(k) f(k, s) \mathrm{d}k$ and matrix $F= \int W(k) f(k, s) f(k, s)^T \mathrm{d}k$, yielding the solution $\hat{\theta} = F^{-1} D$, where $F^{-1}$ also happens to be the uncertainty on these parameters for a fixed $s$ value. You can also reuse most integrals for the computation of $D$ and $F$ by appropriately scaling with $s$, except for the first component involving $f_0(k,s)=P^\mathrm{fid}(k,s)$.} Therefore, the problem of finding the bias reduces almost completely to a simple one-dimensional parameter optimization, for which there exist an abundance of numerical methods. The full minimization in $s$ typically takes less than a second in our code.

\paragraph{Correct by importance sampling.} Because the bias is small, the correction can be applied on the posterior sampling produced by an MCMC at the interpretation step by importance sampling, computing explicitly the relevant bias $\Delta \alpha/\alpha$ and correcting it for every MCMC posterior point (see above for efficient ways of doing this). Another possibility is to correct the $\alpha$ directly as outlined next.

\paragraph{Apply a Taylor approximation of the systematic shift.} We have shown in \cref{sec:results} that the effect is negligible or zero for most parameters beyond $\Lambda$CDM except for $\Omega_{b+\mathrm{cdm}} h^2, N_\mathrm{eff}$ and $\Omega_b h^2$. In particular, we have reviewed compelling arguments that any late-time modification of the cosmological model (after baryon drag) does not alter the bias in $\alpha$ and that early Universe modifications not involving additional radiation, such as modifications to recombination or the introduction of early dark energy, likewise do not introduce additional biases. We therefore consider the following set of parameters:
\begin{equation}\label{eq:parameter_variation_parameters}
    \theta = \{\Omega_{b+\mathrm{cdm}} h^2, N_\mathrm{eff}, \Omega_b h^2\}\,.
\end{equation}

To a very good approximation, the bias depends linearly on most parameter deviations from the fiducial value, except when varying simultaneously two or more of these. In this case, a quadratic expression is used to provide an accurate \textit{prediction} for $\Delta\alpha/\alpha$. We find that third-order corrections are always negligible in the ranges investigated in \cref{sec:results} (see \cref{app:correction_accuracy}). We therefore expand
\begin{equation}
    b(\Delta \theta)\equiv\frac{\Delta\alpha}{\alpha}(\Delta \theta)=
    (\nabla b)^T \Delta \theta +\frac{1}{2} \Delta \theta^T \mathbf{H} \Delta \theta~,
\end{equation}

where $\nabla b$ denotes the Jacobian of the bias $b=\Delta \alpha/\alpha$ and $\mathbf{H} = \nabla \nabla^T b$ denotes the Hessian matrix of the second derivatives. 
Let us define
\[
x \equiv \Delta\Omega_{b+\mathrm{cdm}}h^2, \quad
y \equiv \Delta\Omega_b h^2, \quad
z \equiv \Delta N_{\rm eff}.
\] The Jacobian and Hessian are given in \cref{eq:Jacobian_correction,eq:Hessian_correction}.

\begin{align}
\label{eq:Jacobian_correction}
\mathbf{\nabla b} &=
\begin{pmatrix}
\frac{\partial b}{\partial x} 
\\[1.0em]
\frac{\partial b}{\partial y } 
\\[1.0em]
\frac{\partial b}{\partial z } 
\end{pmatrix}
=
\begin{pmatrix}
 -0.0398 \\[1.0em]
  0.341\\[1.0em]
 -0.00145
\end{pmatrix}~,
\\
\label{eq:Hessian_correction}
\mathbf{H} &=
\begin{pmatrix}
\frac{\partial^2 b}{\partial x^2} & \frac{\partial^2 b}{\partial x \partial y} & \frac{\partial^2 b}{\partial x \partial z} \\[1.0em]
\frac{\partial^2 b}{\partial y \partial x} & \frac{\partial^2 b}{\partial y^2} & \frac{\partial^2 b}{\partial y \partial z} \\[1.0em]
\frac{\partial^2 b}{\partial z \partial x} & \frac{\partial^2 b}{\partial z \partial y} & \frac{\partial^2 b}{\partial z^2}
\end{pmatrix} 
=
\begin{pmatrix}
 0.116   & 0.298   & 0.00305              \\[1.0em]
 0.298   & 0.000       & -0.0126             \\[1.0em]
 0.00305 & -0.0126 & 1.025\times 10^{-4}
\end{pmatrix}~.
\end{align}

\section{Conclusions}\label{sec:conclusions}

In standard BAO analyses of galaxy clustering, cosmological information is encoded in the BAO shifts,  compressed parameters that depend on distances and on the ratio of the sound-horizon scale at radiation drag for the cosmology of interest relative to that of a fiducial cosmology.

The interpretation pipeline (which converts the constraints on the compressed parameters into those of cosmological parameters) typically computes the sound horizon using the defining integral $r_d^{\rm int}$ (\cref{eq:rd_int}). However, the data analysis pipeline derives the sound horizon, $r_d^{\rm obs}$, from the actual measured compressed statistics (power spectrum or correlation function). The difference between the two approaches ($r_d^{\rm int}\not \equiv r_d^{\rm obs}$) introduces a systematic error in cosmological inference and interpretation, which should be quantified and, if non-negligible compared to the statistical errors, accounted for or corrected.

Previous analyses focused on simple extensions of $\Lambda$CDM \cite{Thepsuriya:2014zda,Carter:2019ulk}. This work extends the analysis to more general models, including scenarios that modify the sound horizon in the early universe, and quantifies the associated systematic effects for ongoing state-of-the-art surveys. Although we have presented results for a survey with specifications comparable to DESI, the findings apply to other galaxy redshift surveys such as Euclid or SPHEREx. 

We find that the bias increases approximately linearly with the distance of the cosmological parameters from those of the adopted fiducial model, and it is particularly significant for $\Omega_m$ and $N_{\rm eff}$, where care is required when degeneracies with other beyond $\Lambda$CDM parameters, especially in high-dimensional models, permit large departures from the fiducial parameter values.

For DESI Y1 data, this is not a concern, as the systematic shifts are well below the statistical errors for all reasonable cosmologies. For DESI~Y5 data (and therefore for other stage IV surveys such as Euclid), the effect may become non-negligible. While the parameter variations required to induce a significant bias for $\Omega_b h^2$ are typically excluded from BBN (and/or CMB) constraints, the bias becomes relevant for a shift exceeding  $|\Delta \Omega_m| \approx 0.05$, or  $|\Delta N_\mathrm{eff}| \approx 0.4$. Such large shifts are not ruled out under certain conditions and for certain data sets (see discussion in \cref{ssec:omegam,ssec:neff,ssec:compensated_cosmologies,ssec:special} and for example \cite[Tab.V]{SPT-3G:2025bzu} who report $\Delta N_\mathrm{eff} = 0.47 \pm 0.23$ with SPT-3G+DESI data.) We also propose several approaches to mitigate or correct such biases of varying computational cost: the suitability of each depends on the specific context and application.

This study adopts several simplifying assumptions, most notably, the neglect of the reconstruction procedure and the use of a simplified data covariance matrix. The results of \cite{Carter_2020} suggest that these simplifications do not affect the validity of the findings. 

As this systematic effect may become a limiting factor in error-reduction efforts for current and, in particular, future galaxy surveys, strategies to mitigate or correct for it become crucial. The ones we have outlined are sufficient to reduce the systematics by at least 90\% in the relevant parameter range, with a typical reduction closer to 96\% or higher, depending on the specific parameters considered. We anticipate that these results will be of direct relevance to ongoing and forthcoming dark-energy surveys.

\section*{Acknowledgements}

We thank Adriana Nadal-Matosas for her contributions on the early stages of this work. HGM acknowledges support through the Consolidación Investigadora (CNS2023-144605) of the Spanish Ministry of Science and Innovation and the support of the Ramón y Cajal (RYC-2021-034104). LV, HGM, and FAR acknowledge project PID2022-141125NB-I00 MCIN/AEI and  “Center of Excellence Maria de Maeztu” award to the IC- CUB CEX2024-001451-M funded by MICIU/AEI/10.13039/ 501100011033. NS acknowledges support from the Excellence Cluster ORIGINS which is funded by the Deutsche Forschungsgemeinschaft (DFG, German Research Foundation) under Germany’s Excellence Strategy - EXC-2094 - 390783311, as well as the funding through a Fraunhofer-Schwarzschild Fellowship at the LMU.

\appendix

\section{Idealized case and comparison with previous literature}\label{sec:forecast}

We consider an extremely large and precise survey, \enquote*{HUGE} in \cref{tab:surveys}, in order to investigate the performance of all the $s^{\rm obs}$ in the limit of extremely precise data. 
Neither the \enquote{peak} nor the \enquote{$P(k)$ BAO} are sufficiently accurate in this regime, but we include them for completeness.

\begin{figure}
    \centering
    \includegraphics[width=0.49\linewidth]{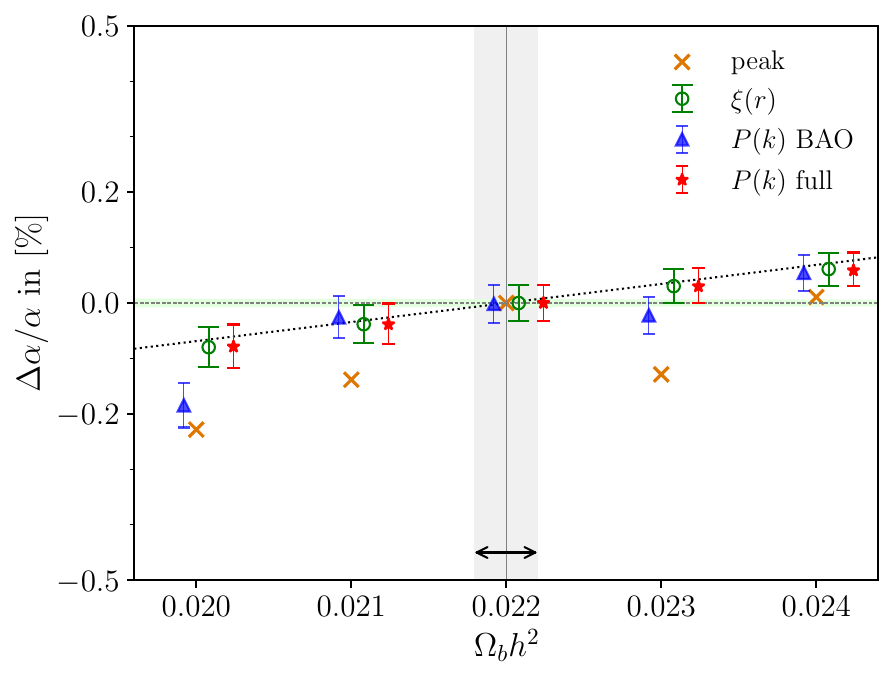}
    \includegraphics[width=0.49\linewidth]{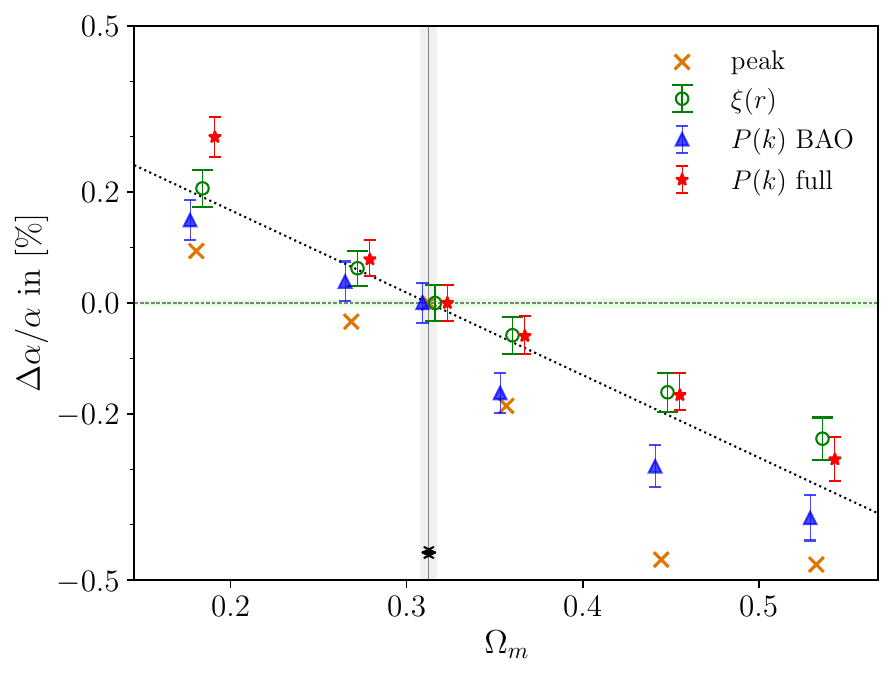}
    \includegraphics[width=0.49\linewidth]{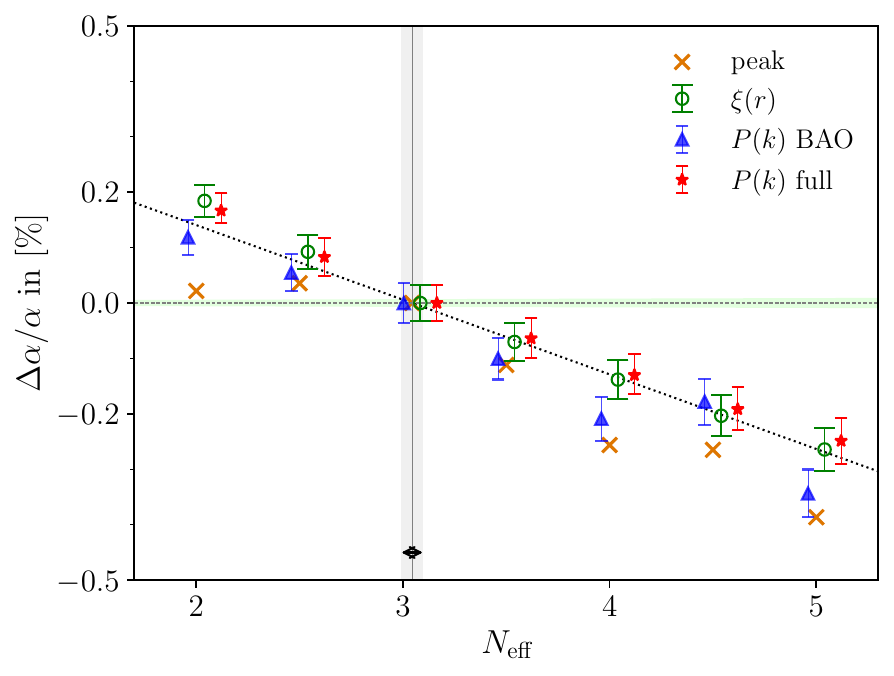}
    \includegraphics[width=0.49\linewidth]{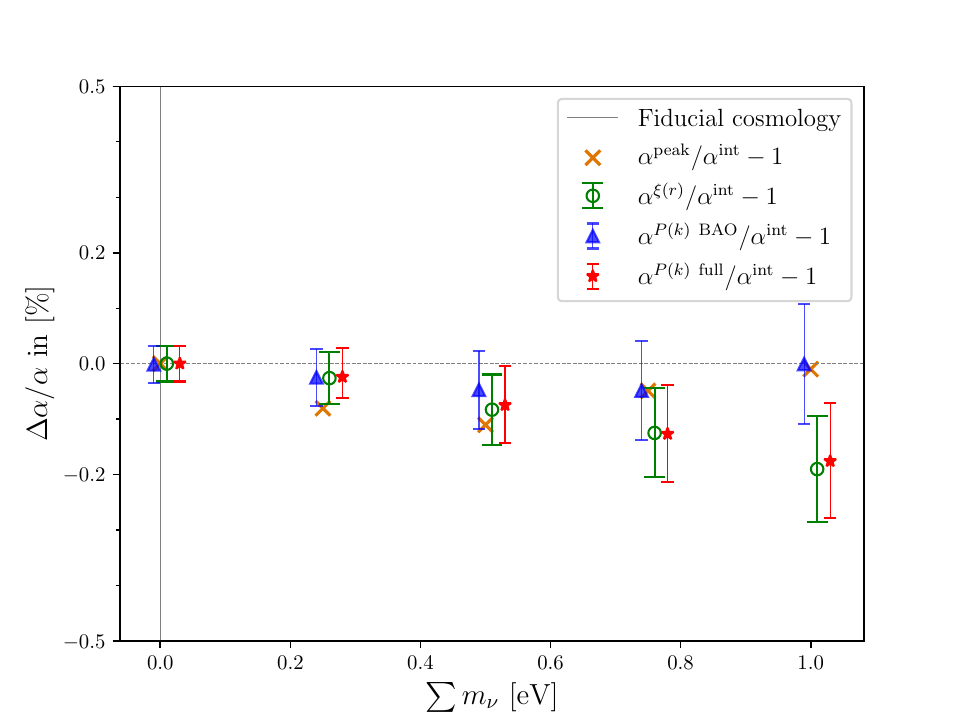}
    \includegraphics[width=0.49\linewidth]{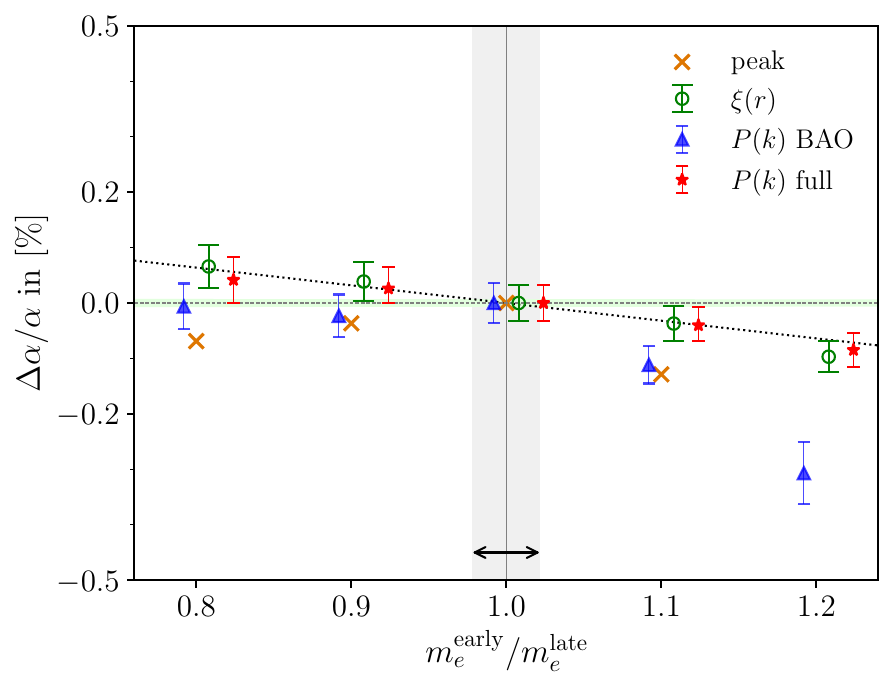}
    \includegraphics[width=0.49\linewidth]{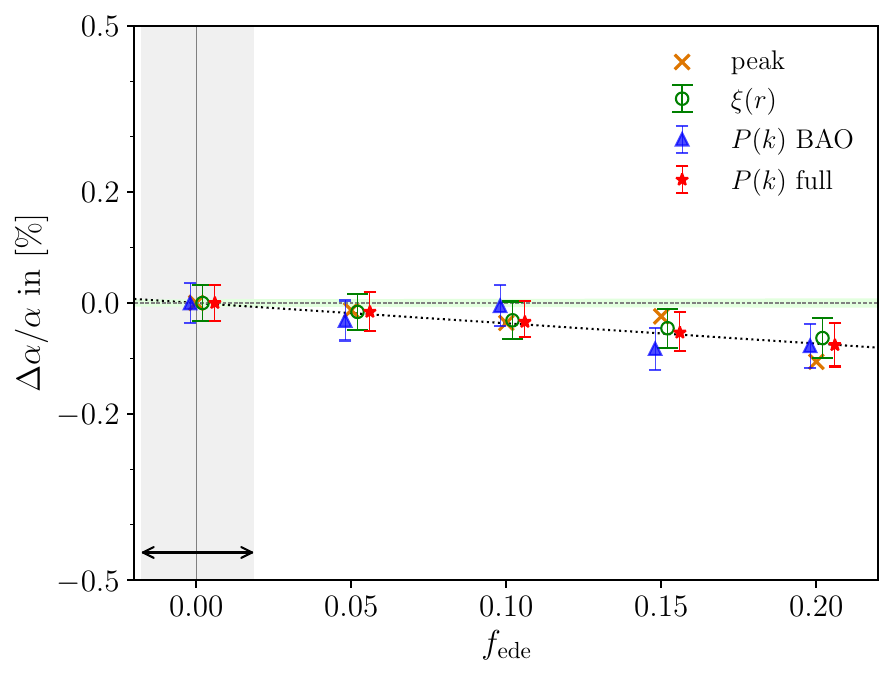}
    \caption{Percentage uncertainty on $\alpha$ for the variation of various cosmological parameters for a survey of extreme precision (\enquote*{HUGE} in \cref{tab:surveys}). }
    \label{fig:extreme}
\end{figure}

We display the results for all cases investigated in \cref{sec:results} in \cref{fig:extreme}. The results of the peak estimation method are evidently not stable at this level of precision, only giving an estimate at the few-per cent level in some cases. Instead, the BAO-only method remains relatively precise but does show permille level differences compared to the most accurate full power spectrum and correlation function methods, which incorporate the additional nuisance parameters to correct for the change in broadband behavior. Our results are broadly consistent with the observations of \cite{Thepsuriya:2014zda}.\footnote{Note that they compare $\Delta r_d/r_d$, while we compare $\Delta \alpha/\alpha \approx -\Delta r_d/r_d$ since $\alpha \propto s \propto 1/r_d$\,.} 

For variations in $N_\mathrm{eff}$, they show in their figure 4 differences at the 0.15\% level, which we confirm.\footnote{We find slightly larger differences beyond 0.2\% for $N_\mathrm{eff}=5$. We have checked that a newer version of \texttt{camb} (v1.3.6) gives a sound horizon integral of 138.76Mpc for $N_\mathrm{eff}=5$ when taking the same fiducial cosmological parameters as in \cref{eq:fiducial} and adjusting just the \texttt{camb} parameter \texttt{num\_nu\_massless} (equivalent to $N_\mathrm{eff}$), while their figure 4 has a value around 143.4Mpc. This motivates us to stipulate that \cite{Thepsuriya:2014zda} likely varied their cosmology up to $N_\mathrm{eff}=4$, which would agree with our \texttt{camb} version in a much more compatible value of 143Mpc -- and would give a more symmetric interval of $|\Delta N_\mathrm{eff}|=1$ around their fiducial value.}

Compared to the variation in \cite{Thepsuriya:2014zda} corresponding to $\Omega_m \in [0.298,0.319]$ we consider a much larger range in values. Since currently different measurements return slightly discrepant values \cite{DESI:2024uvr,Rubin:2023ovl} we feel that such a broader range is justified. In their range, our results indeed only vary by $+0.02\%$ at the lower bound and $-0.01\%$ towards the upper bound, agreeing with what the authors write in section IV about this difference being less than $0.03\%$. As evident from the discussion in \cref{ssec:omegam,ssec:special}, we caution that the larger range we adopted could be relevant for near-future surveys.

For the case of massive neutrinos, we find slightly larger differences (up to $0.2\%$ compared to their $0.15\%$), largely due to the different choice of fiducial model (they adopt the minimal mass of the normal hierarchy in this case as a fiducial, while we remain with the massless fiducial model).

Although somewhat broader in their approach, \cite[Sec.~5]{Carter:2019ulk} also investigates differences in the sound horizon between the data analysis pipeline and the integral formula. In their case, they include additional effects (masking, reconstruction, etc., see \cref{sec:nonlin_compare}) and perform full simulations, but necessarily conclude their argument on a statistical basis. Due to the additional effects that they include, their results are generally comparable but slightly larger in size. For example, they find roughly 0.4\% deviations in $\alpha$ already at $|\Delta N_\mathrm{eff}| = 1$, where we find only 0.2\% deviations. We therefore observe that a potentially good part of the systematics found in \cite{Carter:2019ulk} are explainable by the effects mentioned in this paper, and could at least partially be avoided by changing the theory interpretation pipeline (from computing $r_d^\mathrm{int}$ to $r_d^\xi$ or $r_d^{P(k)~\mathrm{full}}$). Beyond this level of comparison, the methodology of this paper would need to be expanded to account for catalogue-level effects, which are beyond the scope of this work.

\section{Additional results}\label{app:further_results}

For illustrative purposes, we show here some figures that are not relevant enough to be shown in the main body of this work but can help to understand the extent of the impact of the approximation in surveys smaller than DESI~Y5 LRG.

\begin{figure}
    \centering
    \includegraphics[width=0.48\linewidth]{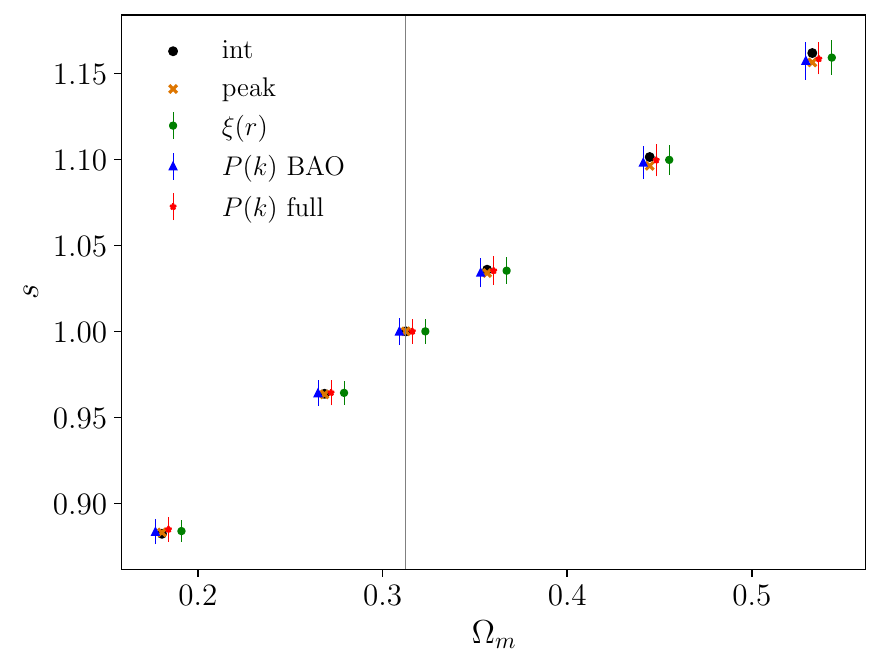}
    \includegraphics[width=0.48\linewidth]{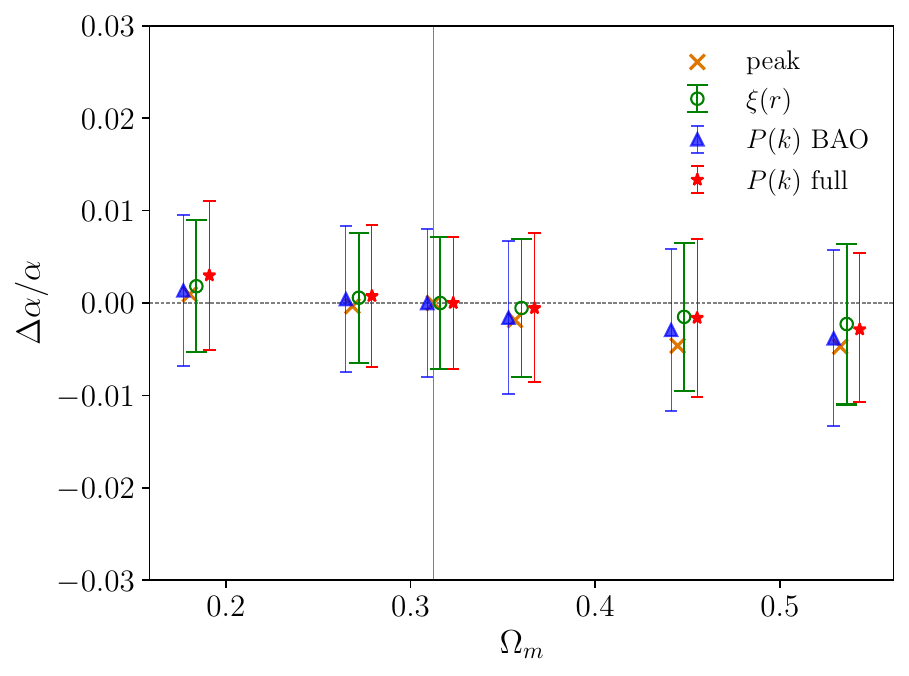}
    \caption{Difference between $\alpha$ from the various methods of \cref{sec:methods} compared to the one obtained from an integral, for the DESI Y1 LRG survey specification for various values of the  $\Omega_m$ parameter.}
    \label{fig:varying_omega_m_desiy1lrg}
\end{figure}

In \cref{fig:varying_omega_m_desiy1lrg}, we see the impact when changing $\Omega_m$ for a DESI~Y1 LRG-like survey. We can appreciate that the values are very well compatible with the integral approximation due to the large error bars.

We also provide additional \cref{tab:bias_analysis_y1,tab:bias_analysis_y5} similar to \cref{tab:bias_analysis} for the individual surveys.

\begin{table}[ht!]
    \centering

    \resizebox{\columnwidth}{!}{

    \begin{tabular}{|c|c c | c c| c c|}
    \hline
        & \multicolumn{2}{|c|}{DESI-Y5 (LRG)} & \multicolumn{2}{|c|}{DESI-Y5 (ELG)} & \multicolumn{2}{|c|}{DESI-Y5 (QSO)} \\
       Parameter  & $\Delta x_-$& $\Delta x_+$  & $\Delta x_-$ & $\Delta x_+$ & $\Delta x_-$ & $\Delta x_+$ \\ \hline \hline
       \rule{0pt}{2.4ex}
$\Omega_b h^2$ & $-0.15366 ^{+ 0.00495}_{- 0.00515}$ & $0.13298 ^{+ 0.00384}_{- 0.00372}$ & $-0.8595 ^{+ 0.0322 }_{- 0.0346}$ & $0.47 ^{+ 0.0102 }_{- 0.00981}$ & $0.86203 ^{+ 0.0029}_{-0.00293}$ & $0.68531 ^{+ 0.00186}_{- 0.00182}$\\
$\Omega_m$ & $-0.033066 ^{+ 0.00269 }_{- 0.00315}$ & $0.033655 ^{+ 0.00327 }_{- 0.00279}$ & $-0.13709 ^{+ 0.00905 }_{- 0.0105}$ & $0.18367 ^{+ 0.0194 }_{- 0.0159}$ & $0.36822 ^{+ 0.00401 }_{- 0.00395}$ & $0.28952 ^{+ 0.00249 }_{- 0.00243}$\\
$N_\mathrm{eff}$ & $-0.34834 ^{+ 0.00882 }_{- 0.00938}$ & $0.37503 ^{+ 0.0109 }_{- 0.0102}$ & $-1.2002 ^{+ 0.0243 }_{- 0.0256}$ & $1.6075 ^{+ 0.0462 }_{- 0.0433}$ & $6.944 ^{+ 0.0388 }_{- 0.0382}$ & $4.7155 ^{+ 0.0178 }_{- 0.0177}$\\
$m^\mathrm{early}_e/m^\mathrm{late}_e$ & $-0.17694 ^{+ 0.0157 }_{- 0.0189}$ & $0.1407 ^{+ 0.0117 }_{- 0.0101}$ & $-0.71762 ^{+ 0.0584 }_{- 0.069}$ & $0.3903 ^{+ 0.0195 }_{- 0.0179}$ & $0.25738 ^{+ 0.000592 }_{- 0.000597}$ & $0.23905 ^{+ 0.000518 }_{- 0.000508}$\\
$f_\mathrm{ede}$ & $-0.12502 ^{+ 0.00271 }_{- 0.00291}$ & $0.14452 ^{+ 0.0039 }_{- 0.0036}$ & $-0.35058 ^{+ 0.00656 }_{- 0.00661}$ & $0.59759 ^{+ 0.0195 }_{- 0.0188}$ & $0.2167 ^{+ 0.0000819}_{- 0.0000809}$ & $0.20979 ^{+ 0.0000758 }_{- 0.0000767}$\\
$\sum m_\nu$ [eV] & $-0.18014 ^{+ 0.00544 }_{- 0.00598}$ & $0.6573 ^{+ 0.0873 }_{- 0.067}$ & $-0.31061 ^{+ 0.00364 }_{- 0.00377}$ & $-0.71225 ^{+ 0.0191 }_{- 0.0199}$ & $-2.9165 ^{+ 0.00901 }_{- 0.00889}$ & $-3.4829 ^{+ 0.0126 }_{- 0.0129}$\\
     \hline
     \end{tabular}
     }
    \caption{Same as \cref{tab:bias_analysis}, but for the separate Y5 surveys of \cref{tab:surveys}.}
    \label{tab:bias_analysis_y5}
\end{table}
\begin{table}[h!]
    \centering

    \resizebox{\columnwidth}{!}{

    \begin{tabular}{|c|c c | c c| c c|}
    \hline
        & \multicolumn{2}{|c|}{DESI-Y1 (LRG)} & \multicolumn{2}{|c|}{DESI-Y1 (ELG)} & \multicolumn{2}{|c|}{DESI-Y1 (QSO)} \\
       Parameter  & $\Delta x_-$& $\Delta x_+$  & $\Delta x_-$ & $\Delta x_+$ & $\Delta x_-$ & $\Delta x_+$ \\ \hline \hline
       \rule{0pt}{2.4ex}
$\Omega_b h^2$ & $-0.57149 ^{+ 0.0204 }_{- 0.0223}$ & $0.36884 ^{+ 0.00917}_{- 0.00859}$ & $40.368 ^{+ 115 }_{- 178}$ & $0.83213 ^{+ 0.0115 }_{- 0.0109}$ & $4.9722 ^{+ 0.026 }_{- 0.0254}$ & $3.5743 ^{+ 0.0133 }_{- 0.0133}$\\
$\Omega_m$ & $-0.10665 ^{+ 0.00828 }_{- 0.00997}$ & $0.11115 ^{+ 0.0109 }_{- 0.00897}$ & $-0.291 ^{+ 0.0154 }_{- 0.0174}$ & $0.82975 ^{+ 0.159 }_{- 0.114}$ & $0.83436 ^{+ 0.0114 }_{- 0.0116}$ & $0.61857 ^{+ 0.00654 }_{- 0.00614}$\\
$N_\mathrm{eff}$ & $-0.98944 ^{+ 0.0229 }_{- 0.0243}$ & $1.2075 ^{+ 0.0364 }_{- 0.0339}$ & $-2.5866 ^{+ 0.0413 }_{- 0.0445}$ & $6.2837 ^{+ 0.27 }_{- 0.238}$ & $17.238 ^{+ 0.135 }_{- 0.135}$ & $10.264 ^{+ 0.0484 }_{- 0.0473}$\\
$m^\mathrm{early}_e/m^\mathrm{late}_e$ & $-0.64014 ^{+ 0.0627 }_{- 0.0784}$ & $0.33302 ^{+ 0.02 }_{- 0.0178}$ & $-4.7727 ^{+ 0.87 }_{- 1.32}$ & $0.77991 ^{+ 0.0286 }_{- 0.0274}$ & $2.7675 ^{+ 0.0303 }_{- 0.0299}$ & $2.0357 ^{+ 0.0165 }_{- 0.0161}$\\
$f_\mathrm{ede}$ & $-0.27055 ^{+ 0.00482 }_{- 0.00504}$ & $0.54018 ^{+ 0.0205 }_{- 0.0189}$ & $-0.6185 ^{+ 0.00784 }_{- 0.00809}$ & $9.6357 ^{+ 2.43 }_{- 1.61}$ & $1.8163 ^{+ 0.00439 }_{- 0.00445}$ & $1.4929 ^{+ 0.00302 }_{- 0.00295}$\\
$\sum m_\nu$ [eV] & $-0.30976 ^{+ 0.00509 }_{- 0.00532}$ & $-1.0571 ^{+ 0.0576 }_{- 0.064}$ & $-0.49084 ^{+ 0.00284 }_{- 0.00289}$ & $-0.68725 ^{+ 0.00558 }_{- 0.00565}$ & $7.4669 ^{+ 0.0417 }_{- 0.0405}$ & $5.7568 ^{+ 0.0243 }_{- 0.0246}$\\
     \hline
     \end{tabular}
     }
    \caption{Same as \cref{tab:bias_analysis}, but for the separate Y1 surveys of \cref{tab:surveys}.}
    \label{tab:bias_analysis_y1}
\end{table}

\begin{table}[ht!]
    \centering

    \resizebox{\columnwidth}{!}{

    \begin{tabular}{|c|c c | c c| c|}
    \hline
        & \multicolumn{2}{|c|}{DESI-Y5 (total)} & \multicolumn{2}{|c|}{DESI-Y1 (total)} & \\
       Parameter  & $\Delta x_-$& $\Delta x_+$  & $\Delta x_-$ & $\Delta x_+$ & Slope of the bias \\ \hline \hline
       \rule{0pt}{2.4ex}
$\Omega_b h^2$ & $-0.39701 ^{+ 0.0157 }_{- 0.0166}$ & $0.23528 ^{+ 0.00574 }_{-0.0056}$ & $-1.7487 ^{+ 0.0901 }_{- 0.0987}$ & $0.64863 ^{+ 0.0131 }_{- 0.0128}$ & $34.5 \pm 1.05$\\
$\Omega_m$ & $-0.070792 ^{+ 0.00575 }_{- 0.00671}$ & $0.074541 ^{+ 0.00748 }_{- 0.00635}$ & $-0.21636 ^{+ 0.0162 }_{- 0.0185}$ & $0.25139 ^{+ 0.0253 }_{- 0.0216}$ & $-0.0149 \pm 0.00131$\\
$N_\mathrm{eff}$ & $-0.7111 ^{+ 0.0203 }_{- 0.0213}$ & $0.84925 ^{+ 0.0305 }_{- 0.0288}$ & $-1.833 ^{+ 0.0371 }_{- 0.0391}$ & $3.0193 ^{+ 0.108 }_{- 0.0994}$ & $-0.00133 \pm 0.0000423$\\
$m^\mathrm{early}_e/m^\mathrm{late}_e$ & $-0.39921 ^{+ 0.0376 }_{- 0.0464}$ & $0.27055 ^{+ 0.0206 }_{- 0.0178}$ & $-2.4358 ^{+ 0.399 }_{- 0.586}$ & $0.55868 ^{+ 0.026 }_{- 0.024}$ & $-0.00323 \pm 0.000268$\\
$f_\mathrm{ede}$ & $-0.24648 ^{+ 0.00509 }_{- 0.0054}$ & $0.33461 ^{+ 0.01 }_{- 0.00931}$ & $-0.61711 ^{+ 0.0114 }_{- 0.012}$ & $1.1271 ^{+ 0.0408 }_{- 0.0376}$& $-0.0037 \pm 0.0000902$\\
$\sum m_\nu$ [eV] & $-0.31196 ^{+ 0.00714 }_{- 0.00758}$ & $-5.3498 ^{+ 1.55 }_{- 3.58}$ & $-0.39147 ^{+ 0.00379 }_{- 0.00403}$ & $-0.67747 ^{+ 0.0117 }_{- 0.0117}$ & $-0.00168 \pm 0.0000834$\\
     \hline
     \end{tabular}
     }
    \caption{Same as \cref{tab:bias_analysis}, but for a threshold for $\Delta x_\pm$ of $\frac{1}{2} \sigma_\alpha$\,.}
    \label{tab:bias_analysis_sig05}
\end{table}

\section{Shot noise calculations}\label{app:shot_noise}
Compressing the information contained within multiple surveys into a single survey is not trivial. Naturally, when the surveys are assumed to be independent, the $\chi^2$ of the individual surveys can be summed. Here, we took a slightly different approach and attempted to find a single higher-precision survey that would act as the analogue of the combination of the individual surveys (LRG, ELG, QSO). For this, \cref{eq:likelihood_full} can be used both for the sums of the individual surveys and the combined survey. They yield respectively
\begin{equation}
    W_1 = \sum V_i / (P^\mathrm{fid}(k) + N_i)^2~, \qquad \qquad W_2= V_\mathrm{comb}/(P^\mathrm{fid}(k) + N_\mathrm{comb})^2
\end{equation}
Naturally, due to the different $k$-dependence on both sides, there is no single solution where $W_1 = W_2$ for all $k$. However, we can use the convexity of the function to make general statements for the weighted power-mean
\begin{equation}\label{eq:N_comb}
    N_\mathrm{comb}^p = \frac{1}{V_\mathrm{comb}} \sum V_i N_i^p
\end{equation}
with $V_\mathrm{comb} = \sum V_i$\,. There are two critical cases of $p$ where either $W_1 > W_2$ for all $k$ or $W_1 < W_2$ everywhere. For this, we look at the function
\begin{equation}
    f(y) = 1/(P^\mathrm{fid}(k)+y^{1/p})^2
\end{equation}
and we note that 
\begin{equation}
    f''(y) = \frac{2 y^{\frac{1}{p}-2} \left[ (p-1) P^{\mathrm{fid}}(k) + (p+2) y^{1/p} \right]}{p^2 \left( P^{\mathrm{fid}}(k) + y^{1/p} \right)^4}
\end{equation}
We observe that if $p>1$ and $y>0$ then $f''(y)>0$ for all $y$. Contrarily, if $p<-2$ and $y>0$ that $f''(y) < 0$ for all $y$. Therefore, by Jensen's inequality, we have
\begin{equation}
    W_2/V_\mathrm{comb} = f\left(\sum w_i N_i^p\right) \geq \sum w_i f\left(N_i^p\right) = W_1/V_\mathrm{comb}
\end{equation}
for $p>1$ (and $\leq$ for $p<-2$), choosing $w_i = V_i / V_\mathrm{comb}$\,. This directly yields $W_1 > W_2$ for all $k$ if $p>1$ and $W_1 < W_2$ for all $k$ if $p<-2$. Note that this happens to also coincides with the shot-noise dominated limit where $N_i \gg P^\mathrm{fid}(k)$ for $p=-2$ and the signal dominated limit where $P^\mathrm{fid} \gg N_i$ for $p=1$. 

To summarize, if $p\geq 1$ ($\sum N_i V_i = N_\mathrm{comb} V_\mathrm{comb}$), we have a conservative approximation of the surveys where it is ensured that the approximation has less constraining power than the summed combination. Instead, if $p\leq -2$ ($\sum V_i/N_i^2 = V_\mathrm{comb}/N_\mathrm{comb}^2$), we have an optimistic approximation of the surveys where it is ensured that the approximation has more constraining power than the summed combination. In order to estimate the maximum possible impact of the bias, we choose the latter case, $p=-2$.

\section{Quantifying the non-linear BAO shift}\label{appendix:quantify_nl_bao_shift}
The values recovered with the non-linear setup of \cref{sec:nonlin_compare} will be shifted by non-linear effects that are not relevant for the mismatch of the sound horizon scale. In this appendix, we clarify how we determine this shift and how we remove it.

We fit the non-linear mock data vectors for different redshifts using the non-linear model with matching cosmology (i.e. we fit non-linear mock data generated with cosmology \textit{x} with a model whose fiducial cosmology is that same cosmology \textit{x}).  The deviation of the recovered $\alpha$ from unity is $\delta \alpha^\mathrm{NL}$. We perform these tests on the DESI Y5 LRG sample, as this effect is expected to be too small for DR1 DESI. The $\delta \alpha^\mathrm{NL}$ so obtained is highly redshift dependent: it vanishes at $z>5$ (where structures are still linear) and grows towards $z=0$, as evident in \cref{tab:non_linearities_with_redshift}. For redshift $z=0.704$, this effect is as large as $\delta\alpha^\mathrm{NL}=0.004$ for the cosmology with low $N_{\rm eff}$ and by $\delta\alpha^\mathrm{NL}=0.003$ for the cosmology with high $N_{\rm eff}$.

\begin{table}
    \centering
    \begin{tabular}{|c|ccc|}
    \hline
        Cosmology                                   & Low $N_{\rm eff}$ (Comp. 2)    & High $N_{\rm eff}$  (Comp. 1) & Fiducial         \\ \hline \hline
        $\delta\alpha^{\rm NL}(z\sim0)$             & $0.006\pm 0.006$               & $0.005\pm 0.010$              & $0.006\pm 0.004$ \\
        $\delta\alpha^{\rm NL}(z=0.704)$            & $0.004\pm 0.005$               & $0.003\pm 0.009$              & $0.004\pm0.003$  \\
        $\delta\alpha^{\rm NL}(z=0.8)$              & $0.003\pm 0.005$               & $0.003\pm 0.009$              & $0.003\pm0.004$  \\
        $\delta\alpha^{\rm NL}(z=5.5)$              & $0.00\pm 0.01$                 & $0.00\pm 0.02$                & $0.00\pm0.03$    \\\hline
    \end{tabular}
    \caption{
    Nonlinear BAO shift $\delta \alpha^{\rm NL}$ as a function of redshift for the fiducial and compensated cosmologies. The reported error bars are statistical for a DESI 5Yr-like survey volume.}
    \label{tab:non_linearities_with_redshift}
\end{table}

The errors in \cref{tab:non_linearities_with_redshift} correspond to the adopted survey specifications and effective volume.  Tests for surveys with \enquote{infinite} precision\footnote{For this case, we consider infinite precision a volume 100 times bigger than the DESI Y5 LRG one.} return consistent $\alpha$ values but with much reduced statistical uncertainties.

We then run the pipeline on the \textit{non-linear} mock data vector and interpret the resulting $\alpha$ shift as $\Delta \alpha^{\rm total}=\Delta\alpha^{\rm non-lin.}+\delta\alpha^{NL}$. The results are reported in \cref{tab:alpha_compensated_desi_pipeline}.

\section{Correction accuracy for n-parameter variations} \label{app:correction_accuracy}

As argued in \cref{sec:correcting}, a Taylor expansion of second order to correct $\Delta\alpha/\alpha$ is accurate for variations of more than just one parameter. In this appendix, we motivate this choice and quantify the difference between the value obtained with the Taylor expansion and the actual value of $\Delta\alpha/\alpha$ obtained in the minimization.

A first-order expansion of the bias $\Delta\alpha/\alpha$ is not accurate when more than one cosmological parameter is varied, with $b$ up to 50\% larger than the actual value. Within the region of interest, this is only the case when $\Omega_{b+\mathrm{cdm}} h^2$ and $N_\mathrm{eff}$ are varied simultaneously, as can be estimated from \cref{eq:Jacobian_correction,eq:Hessian_correction}. We investigate how well a second-order expansion describes $b$ in this scenario. 

Using the second-order Taylor expansion, the largest difference found between the actual value of $\Delta\alpha/\alpha$ and $b$ for all the considered two-parameter-shift combinations of the parameters of \cref{sec:correcting} ($\Omega_{\rm cdm} h^2, \Omega_b h^2, N_{\rm eff}$)) is around $(\Delta\alpha/\alpha - b_{2\mathrm{nd}~\mathrm{order}})/\sigma_{\rm min} = 10\%$, with $\sigma_{\rm min}$ being the error found in the minimization of that specific configuration. The typical error found is $(\overline{\Delta\alpha/\alpha - b_{2\mathrm{nd}~\mathrm{order}}})/\overline{\sigma_{\rm min}} = 4\%$, where the overlines indicate the mean of each of the quantities.

\paragraph{Combination of three parameters} 
Large differences between $\Delta\alpha/\alpha$ and $b$ would indicate the need for higher-order terms in the Taylor expansion. We build eight cosmologies with the largest deviations from fiducial we have considered in \cref{sec:results}, using variations of the three parameters of the subset. We remind the reader that the parameters of reference and the range considered for their variation are $\{\Omega_{\rm cdm} h^2 \in [0.06, 0.22], \Omega_b h^2 \in [0.02, 0.024], N_{\rm eff}\in[2,5]\}$ (with $\Omega_{b+\mathrm{cdm}} h^2= \Omega_\mathrm{cdm}h^2+\Omega_b h^2$). We place one cosmology at each edge of the corresponding hypercube, i.e., we always vary all three parameters simultaneously. The average error from the second-order Taylor expansion is $(\overline{\Delta\alpha/\alpha - b_{2\mathrm{nd}~\mathrm{order}}})/\overline{\sigma_{\rm min}} = 0.5\%$. The worst case yields $(\Delta\alpha/\alpha - b_{2\mathrm{nd}~\mathrm{order}})/\sigma_{\rm min} = 8\%$. This indicates that the second-order Taylor expansion is sufficient.

\bibliographystyle{JHEP}
\bibliography{biblio.bib}

\end{document}